\documentclass[aps,a4paper,showpacs,preprintnumbers,amsmath,amssymb,
  floatfix]{revtex4}
\usepackage{dcolumn}
\usepackage{bm}
\usepackage{amsmath,amssymb}
\usepackage{booktabs,subfigure}
\usepackage{graphicx,psfrag}
\usepackage{epsfig}
\usepackage{color} 
\usepackage{rotating}
\usepackage{axodraw}
\usepackage{slashed}
\usepackage{geometry}
\usepackage{rotating}
\usepackage{subfigure}
\usepackage{caption}
\usepackage{pstricks}
\usepackage{color}
\allowdisplaybreaks[3]
\newcommand{\be}{\begin{equation}}
\newcommand{\ee}{\end{equation}}
\newcommand{\bea}{\begin{eqnarray}}
\newcommand{\eea}{\end{eqnarray}}
\newcommand{\cw}[1][{}]{\ensuremath{\cos^{#1} \theta_{W}}}
\newcommand{\sw}[1][{}]{\ensuremath{\sin^{#1} \theta_{W}}}
\newcommand{\tw}[1][{}]{\ensuremath{\tan^{#1} \theta_{W}}}

\newcommand{\M}{\ensuremath{\mathcal{M}}}

\newcommand{\ETslash}{\ensuremath{E_{\mathrm{T}}\hspace{-1.1em}/\kern0.45em}\xspace}

\numberwithin{equation}{section} 
\usepackage{epsfig}
\usepackage{dcolumn} 
\usepackage{bm} 

\def\gsim{\lower0.5ex\hbox{$\:\buildrel >\over\sim\:$}}
\def\lsim{\lower0.5ex\hbox{$\:\buildrel <\over\sim\:$}}

\begin{document}

\title{
Beam polarization effects in the radiative production of lightest 
neutralinos in $e^+ e^-$ collisions in supersymmetric grand unified models.}

\vskip 2cm

\author{P. N. Pandita$^1$ and
~Monalisa Patra$^2$}

\affiliation{$^1$ Department of Physics, North Eastern Hill University, Shillong 793 002, India \\ %
             $^2$ Centre for High Energy Physics, Indian Institute of Science, Bangalore 560 012, India \\ }

\thispagestyle{myheadings}


\vskip 5cm

\begin{abstract}
\noindent
We study the production of the lightest neutralinos in the  process 
$e^+ e^- \rightarrow \chi_1^0 \chi_1^0 \gamma$ in supersymmetric grand 
unified models for the International Linear  Collider energies with 
longitudinally polarized  beams. We consider cases where the standard model
gauge group is unified into the grand unified gauge groups $SU(5)$, or
$SO(10)$. We have carried out a comprehensive study of this process
in the $SU(5)$ and $SO(10)$ grand unified theories 
which includes the QED radiative corrections.  
We compare and contrast the dependence of the signal cross section
on the grand unified gauge group, and on the different representations of the
grand unified gauge group, when the the electron and positron beams 
are longitudinally polarized. To assess the feasibility of experimentally
observing the radiative production process, we have also considered in detail
the background to this process coming from the radiative neutrino 
production process $e^+ e^- \rightarrow \nu \bar\nu \gamma$ with 
longitudinally polarized  electron and positron beams. In addition we 
have also considered
the supersymmetric background  coming from the radiative production of scalar 
neutrinos in the process $e^+ e^- \rightarrow \tilde \nu \tilde\nu^* \gamma$
with longitudinally polarized beams.
The process  can be a major background
to the radiative production of neutralinos when the scalar neutrinos 
decay invisibly.
\end{abstract}
\maketitle

\section{Introduction}
\label{sec:intro}
In supersymmetric models with $R$ parity~($R_P$) conservation, 
the lightest neutralino
is expected to be the lightest supersymmetric particle~(LSP). Because of 
$R_P$ conservation, the lightest neutralino is absolutely
stable.  Being the LSP, it is the end product of any process that involves 
supersymmetric particles in the final state. Because of its
importance in supersymmetric phenomenology, there have been
extensive studies of the neutralino sector of the minimal supersymmetric 
standard model~(MSSM)~\cite{Nilles:1983ge} and its 
extensions~\cite{Bartl:1989ms, Bartl:1986hp, Pandita:1994ms, 
Pandita:1994vw, Pandita:1994zp, Pandita:1997zt, Choi:2001ww,
Huitu:2003ci, Huitu:2010me}. The discovery of neutralinos is one of the
main goals of present and future accelerators. In particular,
an $e^+ e^- $ collider with a center-of-mass energy of $\sqrt s = 500 $GeV
in the first stage, will be an important tool in determining the parameters
of the underlying supersymmetric model with a high precision~\cite{Aguilar-Saavedra:2001rg,Abe:2001nn,Abe:2001gc, Weiglein:2004hn,Aguilar-Saavedra:2005pw}
The capability of such a linear collider in unravelling the structure 
of supersymmetry (SUSY) can be enhanced by using polarized 
electron and positron beams~\cite{Moortgat-Pick:2005cw}.

When the standard model~(SM) gauge symmetry $SU(2)\times U(1)$ is broken,
the fermionic partners of the two Higgs doublets~($H_1, H_2$) of the MSSM 
mix with the fermionic partners of gauge bosons,  resulting in four neutralino
states $\tilde\chi_i^0, i = 1, 2, 3, 4$, and two chargino states 
$\tilde\chi_j^{\pm}, j = 1, 2.$ The composition and mass of the
lightest neutralino, which depends on the soft 
$SU(2)$ and $U(1)$ gaugino masses, $M_2$ and $M_1,$ on the Higgs(ino)
parameter, $\mu$, and on the ratio  of the  two Higgs  vacuum expectation
values, $\tan\beta \equiv v_2/v_1$, will be crucial for the search for 
supersymmetry at the colliders. The values of the soft gaugino masses
at the electroweak scale depend on the boundary conditions on these 
masses at the grand unified theory (GUT) scale. In most of the studies the gaugino masses 
have been taken to be universal at the GUT scale. However, there is
no particular reason to assume that the soft gaugino masses
are universal at the high scale.  Indeed, it is possible to have
nonuniversal soft gaugino masses in grand unified theories.
We  recall that   soft supersymmetry gaugino masses are generated
from higher-dimensional interaction terms involving gauginos and 
auxiliary parts of chiral superfields~\cite{Cremmer:1982wb}.
For example, in $SU(5)$ grand unified theory, the auxiliary part
of a chiral superfield in higher-dimensional terms can be  
in the representation $\bf{1, 24, 75}$ or $\bf{200}$ or, in general,
some combination of these representations.

When the
auxiliary field of one of the $SU(5)$ nonsinglet chiral superfields
obtains a vacuum expectation value (VEV), then the resulting
gaugino masses are nonuniversal at the grand unification scale. 
Similar conclusions
hold for other supersymmetric grand unified models.
Furthermore, nonuniversal
supersymmetry breaking masses are a
generic  feature in some of the  realistic supersymmetric models.
For example, in anomaly mediated supersymmetry breaking models  
the gaugino masses are not unified \cite{Randall:1998uk, Huitu:2002fg}, 
and hence are not universal.

From the above discussion it is clear, that 
the phenomenology of supersymmetric models depends
crucially on the composition of neutralinos and charginos.  
This in turn depends on the soft gaugino mass parameters
$M_2$ and $M_1$, besides the parameters $\mu$ and $\tan\beta$.
Since most of the models discussed in the literature
assume gaugino  mass universality at the GUT scale,
it is important to investigate the changes in the phenomenology
of broken supersymmetry which results from the 
changes in the composition of neutralinos
and charginos that may arise because of the changes in the pattern
of soft gaugino masses  at the grand unification scale~\cite{Ellis:1985jn}. 
The consequences of nonuniversal gaugino masses at the grand unified scale 
and the resulting  change in boundary conditions 
has been considered in several papers.
This includes the  study of constraints arising from different
experimental measurements~\cite{Barger:1998hp,Huitu:1999vx,Djouadi:2001fa}
and in the study  of
supersymmetric dark matter candidates~\cite{Bertin:2002sq,Corsetti:2000yq}.

Recently in Refs.~\cite{Basu:2007ys, Pandita:2011eh}
a detailed study of the radiative 
production of neutralinos in electron-positron collisions
in low-energy supersymmetric models with
universal gaugino masses at the grand unified scale 
was carried out. Furthermore, we have carried out 
a detailed study of the  radiative production of the lightest neutralinos
in electron-positron collisions in grand unified theories~\cite{Pandita:2012es}.
Since longitudinal beam polarization 
is going to play a crucial role in  electron-positron
collisions, it is important to study its effects  
on the radiative production of the 
lightest neutralinos in  electron-positron colliding
beam experiments in the case of grand unified theories. 

In this paper we shall carry out a detailed study of the 
implications of the nonuniversal
gaugino masses, as they arise in grand unified theories,
for the production of lightest neutralinos in 
electron-positron collisions with longitudinally polarized beams. 
Our purpose is to study the role of  longitudinal beam polarization
as a probe of supersymmetric grand unified theories. For this purpose we shall
consider the case of $SU(5)$ and $SO(10)$ grand unified theories,
these being the typical ones  wherein the standard model can be embedded
in a grand unified gauge group. The motivation of this comes from the
fact that longitudinal beam polarization is a distinct possibility
at the International Linear Collider~(ILC). Studies of this type
have not been carried out so far in the context of grand unified theories.
Since in a large class of models of supersymmetry 
the  lightest neutralino is expected to be the
lightest supersymmetric particle, it will be  one of the first states 
to be produced at the colliders, even if   other SUSY particles
may be  too heavy to be produced.
Moreover, this process is likely to complement the search of the SUSY spectrum at the LHC,
where the squarks and gluinos are likely to be produced and studied in detail.
The radiative neutralino production at the  ILC will, thus, be an
independent study irrespective of whether the  colored sparticles 
are found at the Large Hadron Collider.
A detailed study of this process at the ILC will let us determine the mass
and composition of the lightest neutralino along with its couplings,
which by itself would be an important advance.
The experimental performance of the radiative neutralino production
along with the neutralino mass measurement have been  recently
evaluated in a full detector simulation for the Internaional
Large Detector~\cite{Bartels:2012rg}.
At an electron-positron collider, such as the  
ILC, the lightest neutralino
can be directly produced 
in pairs~\cite{Bartl:1986hp,Ellis:1983er}.
However, it will escape
detection such that the direct production of the lightest
neutralino pair is invisible.
One can, however, look for the signature of 
neutralinos in electron-positron colliders 
in the radiative production process,
\begin{equation}
e^+ + e^-\to\tilde\chi_1^0 + \tilde\chi_1^0 + \gamma.
\label{radiative}
\end{equation}
The signature of this process is  a single high-energy photon with 
missing energy  carried away by the neutralinos. 
In this paper we carry out  a detailed study
of  the process (\ref{radiative}) in supersymmetric
grand unified theories with nonuniversal boundary conditions
at the grand unified scale with polarized electron and positron
beams.  The process~(\ref{radiative}) has been
studied in detail in the minimal supersymmetric  
model ~\cite{Fayet:1982ky,Ellis:1982zz, Grassie:1983kq,
Kobayashi:1984wu,Ware:1984kq,Bento:1985in, Chen:1987ux,Kon:1987gi,
Choi:1999bs, Datta:1996ur, Ambrosanio:1995it}, in  various approximations. 
Calculations have also 
been carried out for the MSSM using general neutralino 
mixing~\cite{ Choi:1999bs, Datta:1996ur, Ambrosanio:1995it}.
This process has also been studied in detail in the 
next-to-minimal supersymmetric model~\cite{Basu:2007ys, Pandita:2011eh}.
On the other hand different large electron positron (LEP)
collaborations~\cite{Heister:2002ut, 
Abdallah:2003np, Achard:2003tx, Abbiendi:2002vz,Abbiendi:2000hh}
have studied the  signature
of radiative neutralino production in detail but have found
no deviations from the SM prediction. Thus, 
they have only been able to set  bounds 
on the masses of supersymmetric particles~\cite{Heister:2002ut,
Abdallah:2003np,Achard:2003tx,Abbiendi:2000hh}.  
Also, the role of longitudinal polarization for process (\ref{radiative}) has
been studied in \cite{Dreiner:2007vm}.

We recall here that
in the SM the radiative neutrino process
\begin{equation}
e^+e^- \to  \nu + \bar\nu + \gamma, 
\label{radiativenu} 
\end{equation}
is the leading process with the same signature as Eq.~(\ref{radiative}). The 
cross section  for the process  (\ref{radiativenu})
depends on the number $N_\nu$ of light neutrino
species~\cite{Gaemers:1978fe}.  This process acts as a main background
to the radiative neutralino production process  (\ref{radiative}).
Furthermore, there is also a  supersymmetric background  to the process (\ref{radiative})
coming from radiative sneutrino production  
\begin{equation}
e^+e^- \to \tilde\nu + \tilde\nu^\ast + \gamma.
\label{radiativesnu}
\end{equation}
We shall consider both these processes, since they form the 
main background to the  radiative process (\ref{radiative}), and are
important for determining the feasibility of observing the
radiative production of lightest neutralinos in electron-positron collisions.

For the signal process (\ref{radiative}), the dominant SM  background 
process (\ref{radiativenu})  proceeds through the exchange of 
$W$ bosons, which couple only to left-handed particles. At the LEP
this dominant background process made it impossible to 
see the possible signal of the radiative process (\ref{radiative}),
even for very light neutralinos.  Furthermore, in the case of
the LHC a search has been made for the final states
in $pp$ collisions, containing a photon~($\gamma$) of large transverse momentum
and missing  energy. These events can be produced by the 
underlying reaction $q \bar q \rightarrow \gamma \chi \bar \chi$, where the
photon is radiated by one of the incoming quarks and where 
$\chi$ is a dark matter candidate~(possibly the lightest neutralino).
The primary background for such a  signal at the LHC is the irreducible
SM background from $Z \gamma \rightarrow \nu \bar \nu \gamma.$
This and other SM backgrounds were taken into account in the LHC analysis.
The observed number of events was found to be in agreement with the
SM expectations for the $\gamma$ + missing  energy events.
From this an upper limit for the production of  $\chi $ in the
$\gamma$ + missing transverse energy state was 
obtained~\cite{Chatrchyan:2012tea}.
In view of these negative results,  the International Linear 
Collider, with the possibility of beam polarization, will be
a good place to look for the process with an energetic photon
and large missing energy in the final state characteristic of 
the reaction (\ref{radiative}). Furthermore, in the case of
the ILC with the possibililty of polarized electron and positron
beams, a suitable choice of beam polarization~($e_R^- e_L^+$)
will significantly reduce the SM  background.


The layout of the paper is as follows.
In Sec.~\ref{sec:exp_cons}, we discuss the constraints on the
supersymmetric particle spectrum arising from the experimental results
from the LHC, the Tevatron, and the LEP.
In Sec.~\ref{sec:composition}, we implement the constraints on the parameter 
space of the grand unified models as they arise from the
constraints on the supersymmetric particle spectrum discussed in 
Sec.~\ref{sec:exp_cons}. 
Here we also calculate the elements of the 
mixing matrix which are relevant for obtaining the couplings of the lightest
neutralino to the electron, selectron, and $Z$ boson
which control  the radiative neutralino 
production process (\ref{radiative}).
We also describe in detail
the typical set of input parameters that is used in our numerical
evaluation of cross sections. The set of parameters that we use
is obtained by imposing various experimental and theoretical
constraints discussed in Sec.~\ref{sec:exp_cons} on the parameter 
space of the minimal supersymmetric standard
model with underlying grand unification. These constraints 
will be used throughout  to arrive at the 
allowed parameter space for different models in this paper. 
Furthermore, in Appendix~\ref{sec:gaugino mass patterns} we briefly
review different patterns
of gaugino masses that arise in grand unified theories. Here 
we will consider grand unified theories based on $SU(5)$ and $SO(10)$ 
gauge groups, and outline the origin of nonuniversal gaugino masses for 
these models.   

In Sec.~\ref{sec:radiative cross section} we summarize the cross section for
the signal process, including the beam polarization and its implications
for the signal cross section. Here we also describe in detail
the effect of QED radiative corrections on the cross section for the
radiative neutralino production cross section.  In Sec.~\ref{sec:numerical}
we evaluate  the cross section for the signal process (\ref{radiative}) in
different grand unified theories with nonuniversal gaugino masses,
using the set of parameters obtained in
Sec.~\ref{sec:composition} for different patterns of gaugino
mass parameters at the grand unified scale.
We have included higher-order QED radiative corrections, as described in
Sec.~\ref{sec:radiative cross section},
in all our calculations.  We  also compare and contrast
the results so obtained  with the
corresponding cross section in the MSSM with universal gaugino masses
at the grand unified scale.  The dependence of the
cross section on the parameters of the neutralino sector, and on the
selectron masses is also studied  in detail.

In Sec.~\ref{sec:backgrounds} we discuss the backgrounds  to
the radiative neutralino production process (\ref{radiative}) from
the SM and supersymmetric processes.  An excess of photons
from radiative neutralino production,
with the longitudinally polarized electron and positron beam,
over the backgrounds measured
through statistical significance is also discussed here and
calculated for different grand unified models.
We summarize our results and conclusions in
Sec.~\ref{sec:conclusions}.


\section{ Experimental Constraints}
\label{sec:exp_cons}

In this section, we discuss the latest constraints on the SUSY particle
spectrum from the  data from the  Large Hadron Collider, 
along with the data from the Tevatron and LEP. At the LHC the search for the 
SUSY particles is carried  through different 
channels and with different final states. The final states can contain jets, 
isolated leptons, and $E^{miss}_T,$ or will have same-sign dileptons or jets
with high $p_T$. Different final states are considered to increase 
the sensitivity to a different SUSY spectrum. Observations at the LHC are in good 
agreement with
the SM expectation; therefore, constraints have been set on the  cross
sections for the SUSY processes. Interpreted differently since no 
supersymmetric 
partners of the SM have been detected lower limits are obtained on their masses.

\subsection{Limits on gaugino mass parameters}

The lightest chargino mass and field content is sensitive to the parameters
$M_2$, $\mu$, and tan$\beta$. At the LEP the search for the lightest chargino
through its pair production has yielded a lower limit on its mass 
\cite{lep-chargino}. The limits obtained depend on the mass of 
the sfermions. For 
the chargino masses  following from nonobservation 
of chargino pair production in $e^+ e^-$ collisions at the LEP,
we have the constraint
\be 
M_{\tilde \chi_1^{\pm}} \gsim 103~~{\rm GeV}.  \label{ch-limit}
\ee 
The limit depends on the sneutrino mass.  For a sneutrino mass below $200$
GeV, the bound becomes weaker, since the production of a chargino pair
becomes more rare due to the destructive interference between $\gamma$ or
$Z$ in the $s$ channel and $\tilde\nu$ in the $t$ channel. In the models we
consider, $m_{\tilde\nu}$ is close to $m_0,$ where $m_0$ is the soft 
SUSY breaking scalar mass.   When $m_{\tilde\nu}<200$
GeV, but $m_{\tilde\nu}>m_{\tilde\chi^\pm}$, the lower limit 
becomes \cite{Yao:2006px}
\be M_{\tilde \chi_1^{\pm}} \gsim 85~~{\rm GeV}.  \ee
For the parameters of the chargino mass matrix, the limit
(\ref{ch-limit}) implies 
an approximate lower limit~\cite{Abdallah:2003xe,Dreiner:2009ic} 
\be M_2,~~ \mu \gsim 100~~{\rm GeV}.
\label{limits1}
\ee 
The lower limits  in Eq.~(\ref{limits1}) on  $M_2$ and $\mu$ are 
obtained by scanning over the MSSM parameter space and are, therefore,
expected to be model independent~\cite{Huitu:2010me}. 
Recently a search was done by the ATLAS experiment 
for the direct production of charginos
and neutralinos in the final states with three leptons and $p_T^{miss}$.
In the context of simplified models degenerate $\tilde{\chi}_1^\pm$ and
$\tilde{\chi}_2^0$ with masses up to $300$ GeV are excluded for large mass
differences with the $\tilde{\chi}_1^0$. For our analyses
we have considered the limit set on the chargino mass from the LEP.
The combination of chargino, slepton and Higgs boson searches has
provided a lower limit on $m_{\tilde{\chi}_1^0}$ as a function of
$\tan\beta$. The absolute lower limit  on the neutralino mass is
$47$ GeV at large $\tan\beta$.

\subsection{Exclusion limits on squarks and gluinos}

The colored SUSY particles, being QCD-mediated processes,can be  more copiously 
produced in the proton-(anti) proton collider with their higher centre-of-mass
energies compared to the LEP. In the context of the Constrained 
Supersymmetric Standard Model (CMSSM),
Tevatron experiments have excluded squark and gluino
masses of $379$ and $308$ GeV, respectively,
based on an integrated luminosity of $2.1$ fb$^{-1}.$

In the framework of the CMSSM,
the LHC experiments with approximately 
5 fb$^{-1}$ of data have excluded gluino
masses below $800$ GeV for all squark masses.
Moreover, squark and gluino masses below approximately 
$1400$ GeV are excluded at
95\% C.L. (for equal squark and gluino masses)~\cite{Atlas_gluino, CMS_gluino}.
The limits, though derived for a 
particular choice of parameters in the context of CMSSM, depend slightly on
the choice. Analyses have also been done setting a limit on gluino mass as a
function of the lightest neutralino. The limits obtained are sensitive 
to the neutralino mass and to the gluino neutralino mass difference.

In the framework of the CMSSM, the LHC experiments have also 
obtained limits on the first- and second-generation squark masses
\cite{Atlas_gluino, CMS_gluino}. They have excluded
masses below around $1300$ GeV for all values of
gluino masses. Similarly,  an analysis is
carried out on the squark mass as a function of the neutralino mass. Overall,
considering all the analyses carried out by LHC in the context of different 
models, 
first- and second-generation squarks along with the gluinos are excluded 
with masses below $1200$ GeV.

The limits on the third-generation squark $\tilde t_1$ mass from LEP is
around $96$ GeV, in
the charm plus neutralino final state~\cite{lep-chargino}. 
Experiments at the LHC and at Tevatron have performed the analyses 
for third-generation squarks in different
scenarios, leading to different final states~\cite{cms-stop}. 
Similar analyses have been carried out for the sbottom
quarks. Overall, for our analyses we will consider the scenario where 
the third-generation squarks are excluded below a mass of about  $800$ GeV.

\subsection{Exclusion limit on slepton masses} 
 
The limits on the selectrons, smuons, and staus masses are from the LEP
experiments~\cite{lep-chargino} because of its clean signature. 
The limits obtained
on the sleptons are sensitive to the lightest neutralino mass. The 
smuons and staus with masses  below $95$ GeV are excluded depending 
on the lightest neutralino $(\tilde{\chi}_1^0)$
mass, provided the mass difference of the slepton 
and $(\tilde{\chi}_1^0)$ is less
than $7$ GeV. A lower limit of around $73$ GeV is set on 
the mass of the right-handed selectron, $m_{\tilde{e}_R},$ independent 
of the neutralino mass.
  
A lower limit of around $45$ GeV is obtained on the sneutrino mass from the 
measurement
of the invisible $Z$ decay width. In the context of the MSSM tighter limits are 
obtained on the mass of sneutrino of around 
$94$~GeV, assuming gaugino mass universality at the GUT
scale.

Taking into account all the constraints set by the different  experiments
as detailed above, for our analyses we have considered 
$m_{\tilde g} \approx$ $1400$ GeV, masses of first two generation squarks
$\approx 1300$ GeV, the third-generation squarks $m_{\tilde {t}, \tilde{b}}$ 
around $1000$ GeV, and the slepton of mass around 
$150$ GeV. For the lightest chargino and neutralino, the LEP limit is respected 
since it gives more stringent bounds compared to the LHC.
The Higgs mass is taken to be consistent with the present LHC results.

\section{Composition of the lightest neutralinos in Grand Unified Theories}
\label{sec:composition}

In this section we list the set of parameters used for our analysis
along with the composition of the lightest neutralino in 
grand unified theories.  In Appendix~\ref{sec:gaugino mass patterns} we review
the patterns  of nonuniversal gaugino masses
in grand unified theories. For the sake of completeness
we have first considered the case of universal gaugino masses in supersymmetric
theories. In Appendix~\ref{neut mass mat} we summarize our notations for the
neutralino mass matrix and the interaction vertices relevant for our study~\cite{Haber:1984rc}.

We have used the set of parameters listed in Table~\ref{parEWSB} for 
our analysis in the case of universal gaugino masses at the grand 
unified scale. The values of the parameters are chosen so as to satisfy 
the various experimental constraints listed in Sec.~\ref{sec:exp_cons}.
We have restricted ourselves to a particular choice of parameter set
with the values of $M_2$ and $\mu$ chosen to correspond to a lightest
neutralino of mass around $108$ GeV. The reason for the choice of this
set was discussed in Ref.~\cite{Pandita:2012es}. 
We call this set of parameters the
MSSM electroweak symmetry breaking  (EWSB) scenario~\cite{Pukhov:2004ca}.
In this scenario we can study the dependence of the neutralino masses as well
as the radiative neutralino production cross section on $\mu$, $M_2$, and the 
selectron masses. 

The composition of the lightest neutralino in case of the MSSM EWSB scenario
for the parameters of Table~\ref{parEWSB} is given by
\begin{eqnarray}
N_{1j} & = & (0.348, -0.175, 0.702, -0.595).
\label{ewsbcomp}
\end{eqnarray}
Thus, the lightest neutralino has a dominant Higgsino component. 
The couplings of the
lightest neutralino to electrons, selectrons, and $Z$ bosons are listed in
table~\ref{feynmandiag} of Appendix~\ref{neut mass mat}. From this Table it
is clear that for a neutralino with  composition (\ref{ewsbcomp}), the 
neutralino - $Z^0$ coupling  is enhanced compared to the
coupling of the lightest neutralino with right and left selectrons 
$\tilde e_{R,L}$.

%
%
\begin{table}
\renewcommand{\arraystretch}{1.0}
\begin{center}
\vspace{0.5cm}
\begin{tabular}{|c|c|c|c|}
\hline
$\tan\beta$ = 10 &$\mu$ = 130 GeV &$M_1$ = 197 GeV &$M_2$ = 395 GeV \\
\hline
$M_3$ = 1402 GeV    &$A_t$= 2800 GeV &$A_b$= 2800 GeV &$A_{\tau}$= 1000 GeV \\
\hline 
$m_{\chi_1^0}$ = 108 GeV &$m_{\chi_1^\pm}$ = 125 GeV 
 &$m_{\tilde e_R}$ = 156.2 GeV &$m_{\tilde\nu_e}$ = 136 GeV \\
\hline
$m_{\chi_2^0}$ = 140 GeV &$m_{\chi_2^\pm}$ = 421.7 GeV  
 &$m_{\tilde e_L}$ = 156.7 GeV &$m_h$ = 125.7 GeV \\ 
\hline
\end{tabular}
\end{center}
\vspace{-0.5cm}
\caption{Input parameters and resulting masses of various 
states in the  MSSM EWSB scenario.}
\renewcommand{\arraystretch}{1.0}
\label{parEWSB}
\end{table}
%

For our analyses, as a benchmark  we have used the radiative neutralino
cross section for the MSSM EWSB scenario with the set of parameters  
as shown in Table~\ref{parEWSB}.

%
\begin{table}[htb]
\renewcommand{\arraystretch}{1.0}
\begin{center}
\vspace{0.5cm}
        \begin{tabular}{|c|c|c|c|}
\hline
 $\tan\beta$ = 10 &$\mu$ = 138 GeV &$M_1$ = 149 GeV &$M_2$ = 890 GeV \\
\hline
 $M_3$ = -2121 GeV    &$A_t$= -1000 GeV &$A_b$= -2700 GeV &$A_{\tau}$= -2700 GeV \\
\hline
$m_{\chi_1^0}$ = 108 GeV &$m_{\chi_1^\pm}$ = 138.7 GeV
 &$m_{\tilde e_R}$ = 156 GeV &$m_{\tilde\nu_e}$ = 136  GeV \\
\hline
$m_{\chi_2^0}$ = 146 GeV &$m_{\chi_2^\pm}$ = 905 GeV
 &$m_{\tilde e_L}$ = 157 GeV &$m_h$ = 124 GeV \\
\hline
\end{tabular}
\end{center}
\vspace{-0.5cm}
\caption{Input parameters and resulting masses for various states in $SU(5)$
supersymmetric grand unified theory with $\Phi$ and $F_\Phi$ in the
{\bf 24}-dimensional representation. We shall refer to this  model
as $[SU(5)]_{24}$ in the text.}
\renewcommand{\arraystretch}{1.0}
\label{parEWSB24}
\end{table}
%
%
\begin{table}[htb]
\renewcommand{\arraystretch}{1.0}
\begin{center}
\vspace{0.5cm}
        \begin{tabular}{|c|c|c|c|}
\hline
 $\tan\beta$ = 10 &$\mu$ = 108 GeV &$M_1$ = -993.9 GeV &$M_2$ = 1172 GeV \\
\hline
 $M_3$ = 1401 GeV    &$A_t$= 1000 GeV &$A_b$= 2700 GeV &$A_{\tau}$= 3000 GeV \\
\hline 
$m_{\chi_1^0}$ = 108 GeV &$m_{\chi_1^\pm}$ = 109 GeV 
 &$m_{\tilde e_R}$ = 156 GeV &$m_{\tilde\nu_e}$ = 136  GeV \\
\hline
$m_{\chi_2^0}$ = 112 GeV &$m_{\chi_2^\pm}$ = 1180 GeV  
 &$m_{\tilde e_L}$ = 157 GeV &$m_h$ = 125 GeV \\ 
\hline
\end{tabular}
\end{center}
\vspace{-0.5cm}
\caption{Input parameters and resulting masses for various states in $SU(5)$
supersymmetric grand unified theory with $\Phi$ and $F_\Phi$ in the
{\bf 75}-dimensional representation. We shall refer to this  model
as $[SU(5)]_{75}$ in the text.}

\renewcommand{\arraystretch}{1.0}
\label{parEWSB75}
\end{table}
%
%
%
\begin{table}[htb]
\renewcommand{\arraystretch}{1.0}
\begin{center}
\vspace{0.5cm}
        \begin{tabular}{|c|c|c|c|}
\hline
 $\tan\beta$ = 10 &$\mu$ = 111 GeV &$M_1$ = 1970 GeV &$M_2$ = 788 GeV \\
\hline
 $M_3$ = 1399 GeV    &$A_t$= 1000 GeV &$A_b$= 2800 GeV &$A_{\tau}$= 3000 GeV \\
\hline 
$m_{\chi_1^0}$ = 107.7 GeV &$m_{\chi_1^\pm}$ = 111 GeV 
 &$m_{\tilde e_R}$ = 166 GeV &$m_{\tilde\nu_e}$ = 136 GeV \\
\hline
$m_{\chi_2^0}$ = 117 GeV &$m_{\chi_2^\pm}$ = 806 GeV  
 &$m_{\tilde e_L}$ = 157 GeV &$m_h$ = 125 GeV \\ 
\hline
\end{tabular}
\end{center}
\vspace{-0.5cm}
\caption{Input parameters and resulting masses for various states in $SU(5)$
supersymmetric grand unified theory with $\Phi$ and $F_\Phi$ in the
{\bf 200}-dimensional representation. We shall refer to this  model
as $[SU(5)]_{200}$ in the text.}
\renewcommand{\arraystretch}{1.0}
\label{parEWSB200}
\end{table}

The input parameters and the resulting masses for the $\bf 24$,
$\bf 75$, and $\bf 200$-dimensional representations of $SU(5)$ 
which result in nonuniversal gaugino masses at the grand unified 
scale obtained in a manner described later in 
Appendix~\ref{subsec:non-universal-gaugino-masses} are shown in 
tables~\ref{parEWSB24}, \ref{parEWSB75} 
and \ref{parEWSB200}, respectively. 
In  arriving at the parameter values in these Tables, 
we have
taken into account various theoretical and phenomenological constraints,
including the electroweak symmetry breaking at the correct scale,
as described  in the Sec.~\ref{sec:exp_cons}. Other values can
be obtained by choosing larger values of the parameter $M_3$.  

The composition of the lightest neutralino for the different 
representations of $SU(5)$ in Table~\ref{tab1} is obtained from the
mixing matrix for the choices of parameters given in Tables~
\ref{parEWSB24}, \ref{parEWSB75} and \ref{parEWSB200}. 
This composition is calculated to be:
\begin{enumerate}
\item
$SU(5)$ with $\Phi$ and $F_\Phi$ in the
{\bf 24}-dimensional representation~(labelled as model $[SU(5)]_{24}$):
\begin{eqnarray}
N_{1j} & = & (0.643,~ -0.053,~ 0.596,~ -0.479);
\label{SU(5)24}
\end{eqnarray}
\item
$SU(5)$ with $\Phi$ and $F_\Phi$ in the
{\bf 75}-dimensional representation~(labelled as model  $[SU(5)]_{75}$):
\begin{eqnarray}
N_{1j} & = & (0.031,~ 0.056,~ -0.712,~ -0.700);
\label{SU(5)75}
\end{eqnarray}
\item
$SU(5)$ with $\Phi$ and $F_\Phi$ in the
{\bf 200}-dimensional representation~(labelled as model  $[SU(5)]_{200}$):
\begin{eqnarray}
N_{1j} & = & (0.018,~ -0.085,~ 0.719,~ -0.689).
\label{SU(5)200}
\end{eqnarray}
\end{enumerate}

\noindent We note from Eqs.~(\ref{SU(5)24}),  (\ref{SU(5)75}),  and (\ref{SU(5)200})
that for the $\bf 24$-dimensional representation of $SU(5)$, the 
dominant component of the neutralino is the bino, 
whereas for the other representations of $SU(5)$, 
there is a Higgsino like lightest neutralino. Thus, 
for $\bf 75$ and $\bf 200$-dimensional representations,
the neutralino, being Higgsino-like, couples weakly to the selectron, 
with the dominant contribution to the cross section coming from the 
neutralino- $Z^0$ coupling.

%
\begin{table}[htb]
\renewcommand{\arraystretch}{1.0}
\begin{center}
\vspace{0.5cm}
        \begin{tabular}{|c|c|c|c|}
\hline
 $\tan\beta$ = 10 &$\mu$ = 116 GeV &$M_1$ = -760 GeV &$M_2$ = 395 GeV \\
\hline
 $M_3$ = 1405 GeV    &$A_t$= 1000 GeV &$A_b$= 2800 GeV &$A_{\tau}$= 3000 GeV \\
\hline 
$m_{\chi_1^0}$ = 108 GeV &$m_{\chi_1^\pm}$ = 111 GeV 
 &$m_{\tilde e_R}$ = 156 GeV &$m_{\tilde\nu_e}$ = 136 GeV \\
\hline
$m_{\chi_2^0}$ = 122 GeV &$m_{\chi_2^\pm}$ = 421 GeV  
 &$m_{\tilde e_L}$ = 157 GeV &$m_h$ = 126 GeV \\ 
\hline
\end{tabular}
\end{center}
\vspace{-0.5cm}
\caption{Input parameters and resulting masses for various states in $SU(5)' \times U(1) \subset SO(10)$
supersymmetric grand unified theory with $\Phi$ and $F_\Phi$ in the
{\bf 210}-dimensional representation with 
$SU(5)' \times U(1)$ in ({\bf 1},{\bf 0}) dimensional representation.
We shall refer to this model as $[SO(10)]_{210}$ in the text.}
\renewcommand{\arraystretch}{1.0}
\label{parSO(10)_210_1}
\end{table}
%
%
\begin{table}[htb]
\renewcommand{\arraystretch}{1.0}
\begin{center}
\vspace{0.5cm}
        \begin{tabular}{|c|c|c|c|}
\hline
 $\tan\beta$ = 10 &$\mu$ = 118 GeV &$M_1$ = 3038 GeV &$M_2$ = 395 GeV \\
\hline
 $M_3$ = 1398 GeV    &$A_t$= 1000 GeV &$A_b$= 2800 GeV &$A_{\tau}$= 3000 GeV \\
\hline 
$m_{\chi_1^0}$ = 108 GeV &$m_{\chi_1^\pm}$ = 113 GeV 
 &$m_{\tilde e_R}$ = 156 GeV &$m_{\tilde\nu_e}$ = 136 GeV \\
\hline
$m_{\chi_2^0}$ = 126 GeV &$m_{\chi_2^\pm}$ = 422 GeV  
 &$m_{\tilde e_L}$ = 157 GeV &$m_h$ = 125.7 GeV \\ 
\hline
\end{tabular}
\end{center}
\vspace{-0.5cm}
\caption{Input parameters and resulting masses for various states in $SU(5)' \times U(1) \subset SO(10)$
supersymmetric grand unified theory with $\Phi$ and $F_\Phi$ in the
{\bf 770}-dimensional representation with $SU(5)' \times U(1)$ in ({\bf 1},{\bf 0})
dimensional representation. We shall refer to this model as $[SO(10)]_{770}$
in the text.}
\renewcommand{\arraystretch}{1.0}
\label{parSO(10)_770_1}
\end{table}
%
\begin{table}[htb]
\renewcommand{\arraystretch}{1.0}
\begin{center}
\vspace{0.5cm}
        \begin{tabular}{|c|c|c|c|}
\hline
 $\tan\beta$ = 10 &$\mu$ = 113 GeV &$M_1$ = 378 GeV &$M_2$ = 985 GeV \\
\hline
 $M_3$ = 1402 GeV    &$A_t$= 1000 GeV &$A_b$= 2800 GeV &$A_{\tau}$= 3000 GeV \\
\hline 
$m_{\chi_1^0}$ = 108 GeV &$m_{\chi_1^\pm}$ = 115 GeV 
 &$m_{\tilde e_R}$ = 156 GeV &$m_{\tilde\nu_e}$ = 136 GeV \\
\hline
$m_{\chi_2^0}$ = 121 GeV &$m_{\chi_2^\pm}$ = 998 GeV  
 &$m_{\tilde e_L}$ = 157 GeV &$m_h$ = 125 GeV \\ 
\hline
\end{tabular}
\end{center}
\vspace{-0.5cm}
\caption{Input parameters and resulting masses for various states in 
$SU(4) \times SU(2)_R \times SU(2)_L \subset SO(10)$
supersymmetric grand unified theory with $\Phi$ and $F_\Phi$ in the
{\bf 770}-dimensional representation with $SU(4) \times SU(2)_R$ in ({\bf 1},{\bf 1})
dimensional representation. We shall refer to this  model 
as $[SO(10)]_{770'}$ in the text.}
\renewcommand{\arraystretch}{1.0}
\label{parSO(10)_770_2}
\end{table}

Similarly in the case of $SO(10)$, for the parameters of 
Tables~\ref{parSO(10)_210_1},  \ref{parSO(10)_770_1}, and
\ref{parSO(10)_770_2}  the composition of the lightest neutralino is given
by the following :
\begin{enumerate}
\item 
$SO(10)$ where $SU(5)' \times U(1) \subset SO(10)$ with $\Phi$ and $F_\Phi$ 
in the {\bf 210}-dimensional representation with 
$SU(5)' \times U(1)$ in ({\bf 1},{\bf 0})-dimensional 
representation~(labelled as model $[SO(10)]_{210}$): 
\begin{eqnarray}
N_{1j} & = & (0.038,~ 0.194,~ -0.719,~ 0.666);  
\label{SO(10))210_1}
\end{eqnarray}
\item 
$SO(10)$ where $SU(5)' \times U(1) \subset SO(10)$ with 
$\Phi$ and $F_\Phi$ in the {\bf 770}-dimensional representation 
with $SU(5)' \times U(1)$ in ({\bf 1},{\bf 0})-dimensional
representation~(labelled as model $[SO(10)]_{770}$): 
\begin{eqnarray}
N_{1j} & = & (0.011,~ -0.193,~ 0.724,~- 0.663);  
\label{SO(10))770_1}
\end{eqnarray}
\item 
$SO(10)$ where $SU(4) \times SU(2)_R \times SU(2)_L \subset SO(10)$ 
with $\Phi$ and $F_\Phi$ in the {\bf 770}-dimensional representation 
with $SU(4) \times SU(2)_R$ in ({\bf 1},{\bf 1})-dimensional
representation~(labelled as model $[SO(10)]_{770'}$): 
\begin{eqnarray}
N_{1j} & = & (0.126,~ -0.660,~ 0.721,~- 0.678),  
\label{SO(10))770_2}
\end{eqnarray}
\end{enumerate}
implying thereby that a Higgsino  is the dominant component for the $\bf 210$- 
and $\bf 770$-dimensional representations with the embedding
$SU(5)' \times U(1) \subset SO(10)$ and for the $\bf 770$-dimensional representation with the
embedding $SU(4) \times SU(2)_R \times SU(2)_L \subset SO(10)$.

Thus, in these cases the dominant contribution to the radiative
neutralino production cross section will come from the 
neutralino-$Z^0$ coupling. Since the LSP for most of the scenarios considered here has a dominant higgsino component, 
the $Z^0$ width imposes a strict constraint, as
the $Z^0$ decay rate involves coupling to the Higgsino component of the neutralino.
We have imposed the LEP constraint on the anomalous $Z^0$ decay width in our calculations :
\begin{equation}
 \Gamma(Z \rightarrow \tilde \chi_1^0 \tilde \chi_1^0) < 3 {\rm MeV}.
\end{equation}

\section{Radiative Neutralino Production in Grand Unified Theories}
\label{sec:radiative cross section}
\begin{figure}[h!]
{\unitlength=1.0 pt
\SetScale{1.0}
\SetWidth{1.0}      
\includegraphics{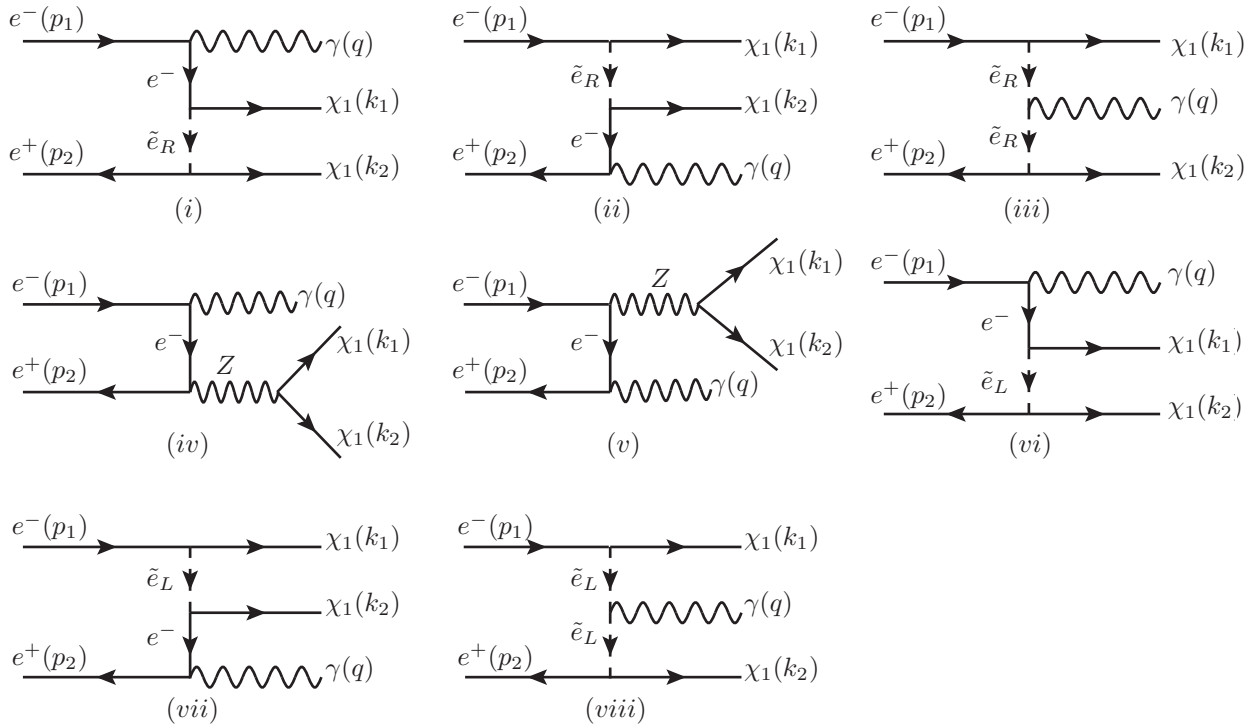}
}
\caption{Feynman diagrams contributing to the  radiative neutralino production 
 $e^+e^- \to  \tilde\chi_1^0\tilde\chi_1^0\gamma.$ There are six other diagrams
which are exchange diagrams corresponding to 
($i$, $ii$, $iii$, $vi$, $vii$, $viii$), with
$u$-channel exchange of selectrons, wherein the neutralinos are crossed
in the final state.}
\label{fig:radneutralino}
\end{figure}

In this section we  calculate the cross section for the  radiative 
neutralino production process 
\begin{eqnarray}
e^-(p_1) + e^+(p_2) \rightarrow \tilde{\chi}_1^0(k_1) + \tilde{\chi}_1^0(k_2)
+ \gamma(q),
\label{radiative1}
\end{eqnarray}
for the case of longitudinally polarized electron and positron beams for 
$SU(5)$ and $SO(10)$ grand unified theories with nonuniversal gaugino
masses at the grand unified scale. The four-momenta of the corresponding
particles are shown by the symbols in the brackets. We show in Fig.~\ref{fig:radneutralino}
the Feynman diagrams contributing to the radiative neutralino production at the 
tree level. The neutralino mixing matrix (\ref{mssmneut}) summarized in 
Appendix~\ref{neut mass mat} determines the couplings of the neutralinos to 
electrons, the selectrons, and to the $Z^0$ bosons. The respective values of 
the soft SUSY breaking gaugino mass parameters $M_1$ and $M_2$ 
for different grand unified models 
have been calculated in Appendix~\ref{sec:gaugino mass patterns}. 
We further note that the elements of the neutralino mixing matrix $N_{1j}$
for the different models considered here, were calculated in the previous section.




\noindent

\subsection{Cross section for the signal process}

At the tree level the process (\ref{radiative1}) proceeds
via the $t$- and $u$-channel exchange of right and left selectrons
$\tilde e_{R,L}$  and via $Z$ boson exchange in the $s$ channel
for the different scenarios considered here as can be seen from
Fig.~\ref{fig:radneutralino}. 
The differential cross section for the process (\ref{radiative1}) 
can be written as~\cite{Grassie:1983kq, Eidelman:2004wy}
\begin{eqnarray}
d \sigma &=& \frac{1}{2} \frac{(2\pi)^4}{2 s}
\prod_f \frac{d^3 \mathbf{p}_f}{(2\pi)^3 2E_f}\delta^{(4)}(p_1 +
p_2 - k_1 - k_2 - q)|\M|^2,
\label{phasespace}
\end{eqnarray}
where $\mathbf{p}_f$ and $E_f$ are the final three-momenta
$\mathbf{k}_1$, $\mathbf{k}_2$, $\mathbf{q}$
and the final energies
$E_{\chi_1}$, $E_{\chi_2}$, and $E_\gamma$
of the neutralinos and the photon, respectively.
Using the standard technique, we sum over the spins of the neutralinos
and the polarization of the outgoing photon. 
The squared matrix
element $|\M|^2$ in Eq.~(\ref{phasespace}) can then be written 
as~\cite{Grassie:1983kq}
\begin{eqnarray}
|\M|^2 & = & \sum_{i \leq j} T_{ij}, \label{squaredmatrix}
\end{eqnarray}
where $T_{ij}$ are squared amplitudes corresponding to the Feynman diagrams
in  Fig.~\ref{fig:radneutralino}. 
The phase space for the radiative neutralino production process 
in  Eq.~(\ref{phasespace}) is described in detail in Ref.~~\cite{Grassie:1983kq}.

\subsection{Longitudinal beam polarization} \label{subsec:LP}

At the future linear collider, the use of beam polarization will significantly
benefit the physics program.  In the case of many processes, it is found that a suitable 
choice of beam polarizations can enhance the signal and suppress the background. 
At the ILC, a beam polarization of $\geq$ 80\% for electrons and $\geq$ 30\%
for positrons at the interaction point is proposed, with a possible upgrading
to about 60\% for the positron beam. In the case of an electron and positron
beam with arbitrary degree of longitudinal beam 
polarization, the total cross section in the centre-of-mass frame with center-of-mass 
energy $\sqrt{s}$ is given by
\begin{eqnarray}
\sigma_{P_{e^{-}}P_{e^{+}}} &=&  \frac{1}{4}\left[(1+P_{e^{-}})(1-P_{e^{+}})\sigma_{RL}
    +(1-P_{e^{-}})(1+P_{e^{+}}) \sigma_{LR}\right]
    \label{eq:dsigPl}
\end{eqnarray}

\noindent In  Eq.~(\ref{eq:dsigPl}) the dependence of the cross section on  
the polarization is parametrized through the degree of polarization, which is
defined as $P_{e^\mp}=(N_R-N_L)/(N_R+N_L)$, where $N_{L,R}$ denote the number 
of left-polarized and right-polarized electrons (or positrons) respectively. 
Moreover $\sigma_{RL}$ denotes the cross section when the electron beam is
completely right  polarized with $P_{e^-}$ = 1,  and the positron beam is 
completely left polarized with $P_{e^+}$ = -1. An analogous definition
holds for $\sigma_{LR}$. We do not take into account the helicity combinations
for the cross section ($LL$ and $RR$) as they are absent in the SM and for the
supersymmetric process considered here. For the signal process,
the significant contribution comes from the selectron or $Z$ exchange
depending on the composition of the neutralino. For all the scenarios
considered, the neutralino is dominantly a Higgsino  with the $Z$ boson exchange
dominantly contributing to the neutralino production process. 
In the case of $SU(5)_{24}$,
the neutralino has a significant bino component resulting in significantly
larger coupling to right selectron; therefore, the production process proceeds
mainly via the exchange of right selectron $\tilde{e}_R$. On the other hand, the
SM background radiative neutrino process proceeds mainly through the exchange of
$W$ bosons, which couple only to left handed particles. Therefore, a polarization
combination with positive electron beam polarization and negative positron beam 
polarization will significantly reduce the background and increase the signal
for the cases where the neutralino has a dominant bino component. When the neutralino
is of a Higgsino type, there is no appreciable change in cross section for this choice 
of beam polarization as $Z$ couples to both left- and right-handed fermions. Since
with this particular choice of beam polarization the SM background decreases, we
present our result for this case with electron beam polarization $P_{e^-}$ = 0.8
and positron beam polarization $P_{e^+}$ = -0.6 as planned for the future linear
collider.

\subsection{Radiative corrections}
\label{sec:radiative corrections}

The future high-energy $e^+e^-$ colliders, in order to avoid energy losses from
synchrotron radiation, are designed as linear colliders. These colliders will
achieve high luminosity through beams with bunches of high number densities. 
Although the high density of charged particles increases the machine luminosity,
it also leads to the generation of a strong electromagnetic field in and around
every colliding bunch. Initial state radiation (ISR), also known as bremsstrahlung,
which results from the interaction of the beam constituents with the accelerating
field, is the most important QED correction to the Born cross section. Along with it,
the interaction of the beam constituents due to the strong magnetic field generated
by the other beam also results in radiation and is known as the beamstrahlung
phenomenon. The general feature of both these cases results in multiple emissions
of photons, both soft and hard, which not only reduces the initial beam energy but also 
results in the disturbance of the initial beam calibration. Moreover, at higher
energies these radiative effects result in messier backgrounds with the 
radiated photons leading to the production of lepton pairs and hadrons. 
The resulting spectrum of the electrons due to the ISR effects mainly depends on
the electron or positron beam energy and the reduced momentum of the incoming 
electron or positron. The photon radiation takes into account the missing 
momentum. On the other hand, the resulting spectrum due to beamstrahlung,
apart from depending on the beam energy, is mainly machine
specific depending on the number of electrons and positrons in a bunch $N_e$, 
the transverse bunch sizes $\sigma_x, \sigma_y$ and the bunch length $\sigma_z$.
Therefore most future machine designs try to minimize the radiation effects by
adjusting the parameters of the bunches accordingly. 

Apart from being a serious problem, 
the radiated photons have also been used in the study
of new physics. The majority of the emitted photons are soft and are lost down
the beam pipe. Only the hard photons with large transverse momentum can be 
tagged, and earlier they were used by the LEP experiments to look for the
invisible final states. The most famous example is the neutrino counting
process $e^+e^-\rightarrow\gamma\nu_l\bar{\nu}_l$ in the standard model, with the
final state being a single photon and large missing energy. This search with a hard
photon tag is similar to the supersymmetric process considered here in our work.
In the case of LEP running at energies beyond the $Z$ resonance, these radiative
effects lead to ``return of the $Z$ peak'' causing a hugely increased cross section.
This was mainly due to the multiple emissions of photons resulting in the electron 
positron pair returning to the $Z$ resonance. Therefore, taking into account all the
above facts, the effect of the radiative effects, both ISR and beamstrahlung, is crucial
for most experimental analyses.

Several strategies exist to include the radiative corrections in the calculations
which have been studied exclusively in the past~\cite{Kureav, Nicorsini,Skrzypek:1990qs,Blankenbecler:1987rg}
in the context of the future linear colliders. We have calculated the radiative
effects for our process and the background processes using 
CalcHEP~\cite{Pukhov:2004ca}, with parameters given in Table~\ref{ilc_par}~\cite{Brau:2007zza}.
In CalcHEP the energy spectrum of the electron and positron is calculated by using 
the structure function formalism. The main idea here is to include the 
radiative corrections by a probability density to find an electron with reduced
momentum inside an incoming electron. This is quite similar to the techniques 
adopted for the hadronic interactions. The total cross section is defined as 
the leading-order cross section convoluted with the structure functions 
including radiative effects. These structure function of the initial leptons 
are valid up to all orders in perturbation theory.  
We emphasize that in this paper 
the radiative effects are included in all our calculations
of the signal and background processes.

\begin{table}[htb]
\renewcommand{\arraystretch}{1.0}
\begin{center}
\vspace{0.5cm}
\begin{tabular}{||c|c||}
\hline 
Collider parameters &ILC \\ \hline
$\sigma_x$ (nm) & 640  \\ 
 &\\
$\sigma_y$ (nm) & 5.7 \\  
&\\
$\sigma_z$ ($\mu$m) & 300 \\  
& \\
N ($10^{10}$) &2 \\  \hline
\end{tabular}
\end{center}
\caption{Beam parameters for the ILC, where N is the number of particles in the bunch and $\sigma_x$, $\sigma_y$
are the transverse bunch sizes at the interaction point, with $\sigma_z$ as the bunch length.}
\label{ilc_par}
\end{table}

\bigskip
\section{Numerical Results}
\label{sec:numerical}

We have calculated the tree-level cross section for radiative neutralino
production (\ref{radiative1}),  the standard model background from 
radiative neutrino production (\ref{radiativenu}), and the supersymmetric
background from sneutrino production (\ref{radiativesnu})
with longitudinally polarized electron and positron beams
using the program  CalcHEP~\cite{Pukhov:2004ca}. 
As noted above we have included the effects of radiative corrections
to the signal as well as the background processes.
Due to the emission of soft photons, the tree-level cross sections have 
infrared and collinear divergences. These divergences are
regularized by imposing cuts on the fraction of beam energy carried 
by the photon and the scattering angle of the photon~\cite{Grassie:1983kq}. 
We define the fraction of the beam energy carried by the photon as $ x = E_{\gamma}/E_{\rm beam},$
where  $ \sqrt {s} = 2E_{\rm beam}$ is the center-of-mass energy, and 
$E_{\gamma}$ is the energy carried away by the photon. 
The following cuts are then imposed on $x$  and on the scattering 
angle $\theta_{\gamma}$ of the photon~\cite{Dreiner:2006sb}:
\begin{eqnarray}
0.02 & \le & x \le  1-\frac{m_{\chi_1^0}^2}{E_{\rm beam}^2}, \label{cut1} \\
-0.95 & \le & \cos\theta_\gamma \le 0.95.  \label{cut2}
\end{eqnarray}

\noindent The lower and upper cut, Eq. (\ref{cut1}), on the energy of the photon is
a function of the beam energy. Interpreted in a different way, the upper
cut corresponds to the kinematical limit of the radiative neutralino production
process. In order to enhance the signal over the main SM background, with
the neutrinos preferably emitted in the forward direction, the required 
detector acceptance cut, Eq.~(\ref{cut2}), on the photon is applied. Except for 
the cuts on energy and the angular spread, no other cut is found to significantly
reduce the background. Therefore, we have implemented these cuts for both
signal and  background processes in the case of all the scenarios which we have
considered in this work.

\begin{figure}[ht]
\begin{minipage}[b]{0.45\linewidth}
\vspace*{0.45cm}
\centering
\includegraphics[width=6.5cm, height=5cm]{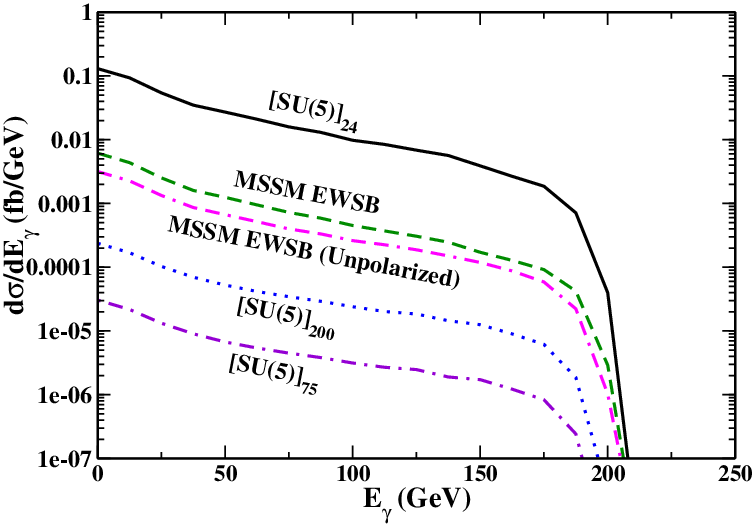}
\caption{The photon energy distribution 
$d\sigma/dE_{\gamma}$ for the radiative neutralino production 
including radiative effects with $(P_{e^-},P_{e^+})$ = 
(0.8, -0.6) in the case of $SU(5)$ with nonuniversal gaugino masses
and in the case of the MSSM~EWSB with universal gaugino masses. 
For comparison we have also shown the case of the MSSM~EWSB with 
unpolarized beams.}
\label{fig:SU(5)}
\end{minipage}
\hspace{0.7cm}
\begin{minipage}[b]{0.45\linewidth}
\centering
\includegraphics[width=6.5cm, height=5cm]{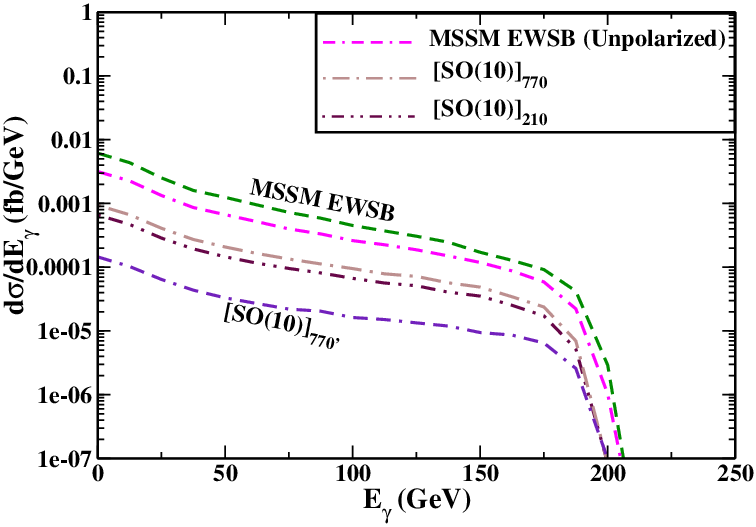}
\caption{The photon energy distribution 
$d\sigma/dE_{\gamma}$ including radiative effects for the radiative 
neutralino production with $(P_{e^-},P_{e^+})$ = (0.8, -0.6) 
in the case of $SO(10)$
with nonuniversal gaugino masses 
and in the case of the MSSM~EWSB with universal gaugino masses.
For comparison we have also shown the case of MSSM~EWSB with
unpolarized beams.}
\label{fig:SO(10)}
\end{minipage}
\end{figure}
    
\subsection{Photon energy~($E_\gamma$) distribution and total beam
energy~($\sqrt {s}$) dependence}
\label{sec:photon_energ_distribution}

First of all we have calculated the energy distribution of the photons 
from the radiative neutralino production in case of the MSSM EWSB and different
GUT scenarios with nonuniversal gaugino mass in the case of 
longitudinal beam polarization.  


\begin{figure}[htb]
\begin{minipage}[b]{0.45\linewidth}
\centering
\vspace*{0.45cm} 
\includegraphics[width=6.5cm, height=4cm]{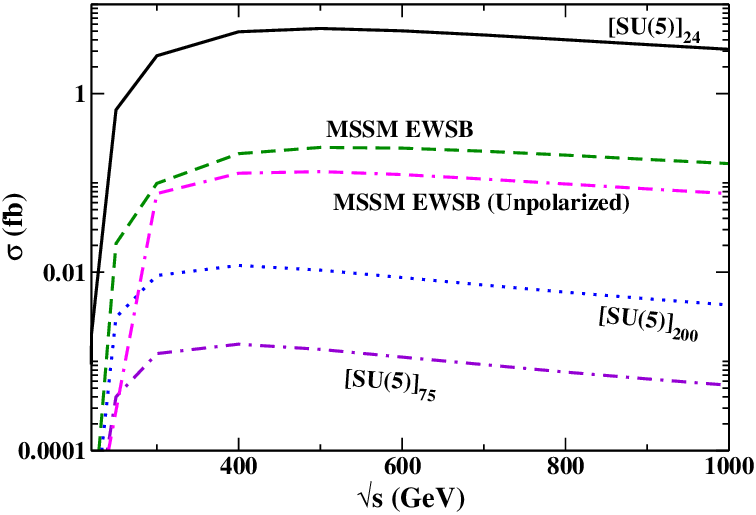}
\caption{Total cross section  $\sigma$ for the signal process,
with the inclusion of 
radiative effects as a function of $\sqrt s$ with $(P_{e^-},P_{e^+})$ = 
(0.8, -0.6) for $SU(5)$ with nonuniversal gaugino masses  and
for the MSSM~EWSB scenario with universal gaugino masses at the grand 
unified scale. For comparison we have also shown the case of MSSM~EWSB 
with unpolarized beams.}
\label{fig:SU(5)_cs}
\end{minipage}
\hspace{0.7cm}
\begin{minipage}[b]{0.45\linewidth}
\centering
\includegraphics[width=6.5cm, height=4cm]{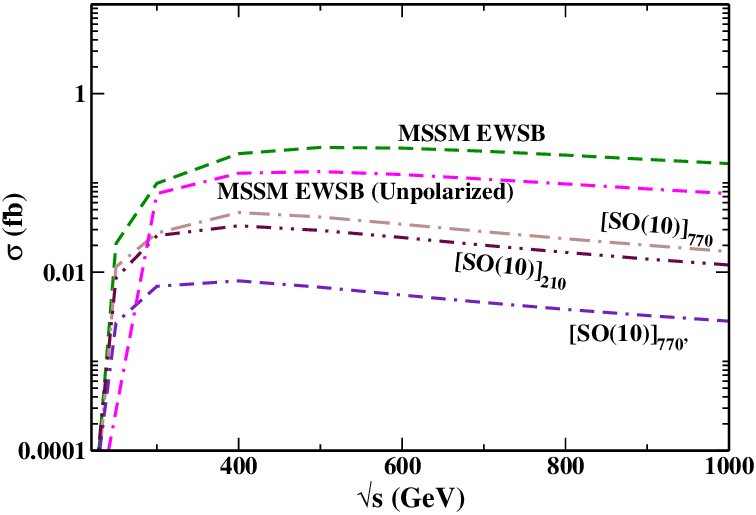}
\caption{Total cross section  $\sigma$ for the signal process,
with the inclusion of 
radiative effects as a function of $\sqrt s$ with $(P_{e^-},P_{e^+})$ = 
(0.8, -0.6) for $SO(10)$ with nonuniversal gaugino masses and
for the MSSM~EWSB scenario with universal gaugino masses at the 
grand unified scale.
For comparison  we have also shown the case of MSSM~EWSB with
unpolarized beams.}
\label{fig:SO(10)_cs}
\end{minipage}
\end{figure} 

The energy distribution of the radiated photon in the presence of longitudinally
polarized beams is shown in Figs.~\ref{fig:SU(5)} and \ref{fig:SO(10)}
for the scenarios with nonuniversal gaugino masses in grand unified 
theories based on  $SU(5)$ and $SO(10)$. 
In these figures the resulting distributions
are also compared with the MSSM EWSB model with universal gaugino masses
at the GUT scale. Similarly the energy dependence of the total cross section
is also calculated with the initially polarized beams and is shown in 
Figs.~\ref{fig:SU(5)_cs} and \ref{fig:SO(10)_cs}. Note that we have included
radiative corrections in all these calculations.  As discussed
before we have restricted ourselves to only right-handed electron beams and left-handed
positron beams in order to reduce the background. The degree of polarization used 
in our calculation is $(P_{e^-},P_{e^+})$ = (0.8, -0.6). The unpolarized 
case in case of the MSSM EWSB is also shown in these figures for the sake of comparison.

The signal in the case of MSSM EWSB and $[SU(5)]_{24}$ is enhanced in the polarized
case compared to the other models considered here. The dominant component 
of the neutralino in $[SU(5)]_{24}$ is a bino, whereas in other cases 
the lightest neutralino is dominantly a Higgsino state. The MSSM EWSB scenario
predicts a lightest neutralino with a dominant Higgsino component, but it also
has a significant bino component leading to the enhancement of right 
selectron-electron-neutralino coupling. Therefore the choice of this particular
polarization leads to an increase in the production cross section.
For the other cases with a Higgsino-like neutralino the $t$- and $u$- channel
exchange of $\tilde{e}_{R,L}$ is suppressed, with the only contribution 
coming from off-shell $Z$ decay. The $Z$ boson due to its ability to 
combine with both left- and right-handed fermions does not result in significant
changes with the inclusion of the beam polarization.

\subsection{Dependence on $\mu$ and $M_2$}

Since the mass of the lightest neutralino depends 
on the parameters $\mu$ and $M_2$,  it is important to
study the dependence of cross section for the signal process
on these parameters.  The dependence of the signal cross section
is considered independently on the parameters $\mu$ and $M_2$.
The values of the parameters $\mu$ and $M_2$ are chosen 
in order to avoid color and charge breaking minima, unbounded from below
constraint on scalar potential, and also to satisfy phenomenological constraints 
on different sparticle masses as discussed in Sec.~\ref{sec:exp_cons}.

We have carried out a check on the parameter space used
in our calculations on whether the complete scalar potential has charge
and color breaking minima, which are lower than the electroweak 
symmetry breaking minimum.
The condition of whether the scalar potential is unbounded 
from below has also been checked by us. 
The criteria used for these conditions are
\bea
A_f^2 & < & 3(m_{\tilde f_L}^2 + m_{\tilde f_R}^2 + \mu^2 + m_{H_2}),
\label{ccb1}\\
m_{H_2} +  m_{H_1} & \ge & 2|B\mu|,
\label{ufb1}
\eea
respectively, at a scale $Q^2 > M^2_{\rm EWSB}$. Here $f$ denotes the
fermion generation, and A is the trilinear supersymmetry breaking parameter.
We have implemented these conditions through the 
SuSpect package~\cite{Djouadi:2002ze} which computes
the masses and couplings of the supersymmetric partners of the SM particles.
For each model considered in this paper, we perform the renormalization group
evolution to calculate the particle spectrum. While doing so we check 
for the consistency of the chosen parameter set
with electroweak symmetry breaking and also whether the conditions (\ref{ccb1}) and
(\ref{ufb1}) are satisfied.

\begin{figure}[htb]
\begin{minipage}[b]{0.45\linewidth}
\centering
\vspace*{0.7cm}
\includegraphics[width=6.5cm, height=5cm]{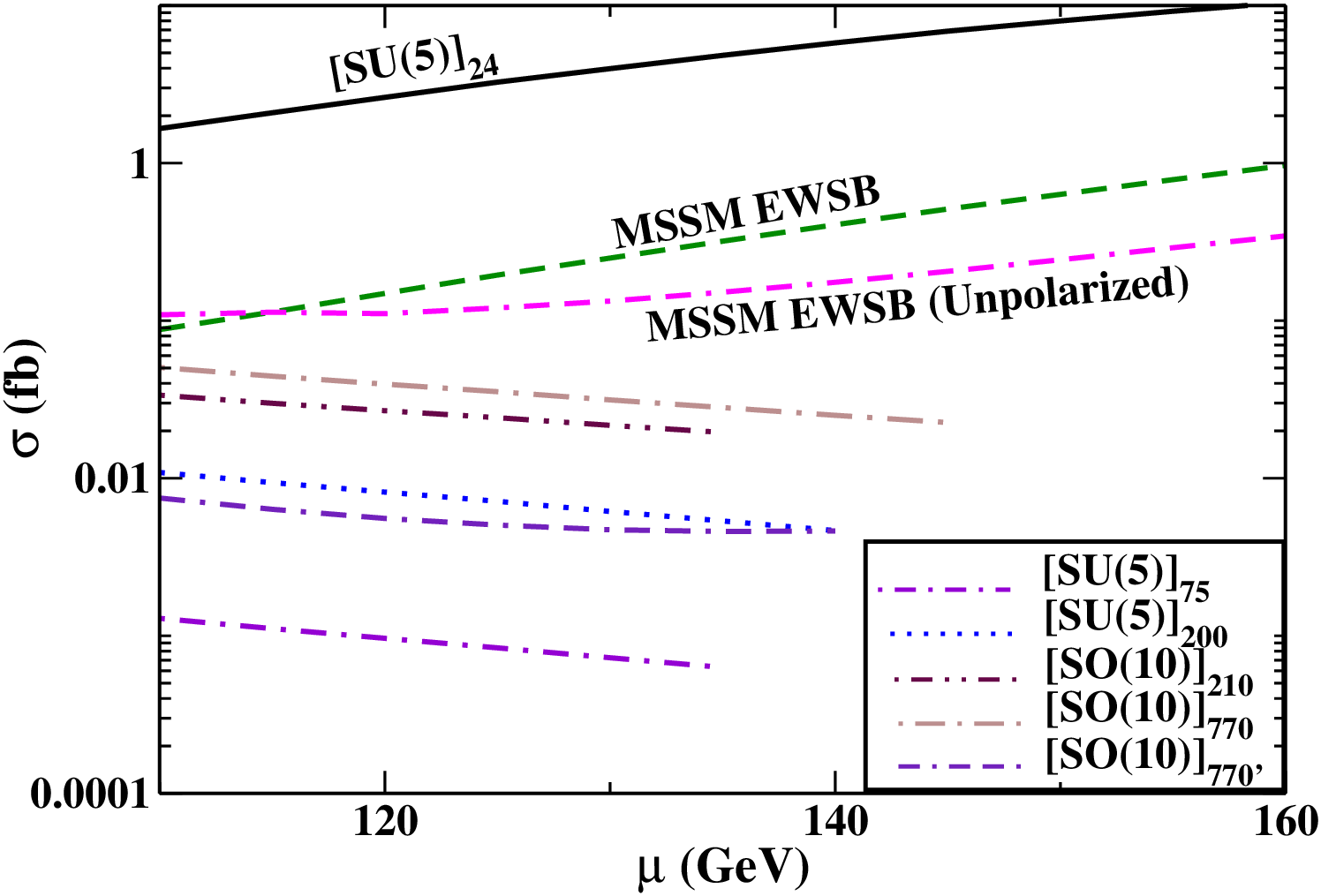}
\caption{The total  radiative neutralino production cross section $\sigma$ with
radiative effects included as a function of $\mu$  in the range 
$\mu$ {\Large{$\epsilon$}} [110, 160] GeV for different models
at $\sqrt{s} = 500$~GeV with $(P_{e^-},P_{e^+})$ = (0.8, -0.6).
For comparison we have also shown the case of MSSM~EWSB with
unpolarized beams.}
\label{fig:mu}
\end{minipage}
\hspace{0.4cm}
\begin{minipage}[b]{0.45\linewidth}
\centering
\includegraphics[width=6.5cm, height=5cm]{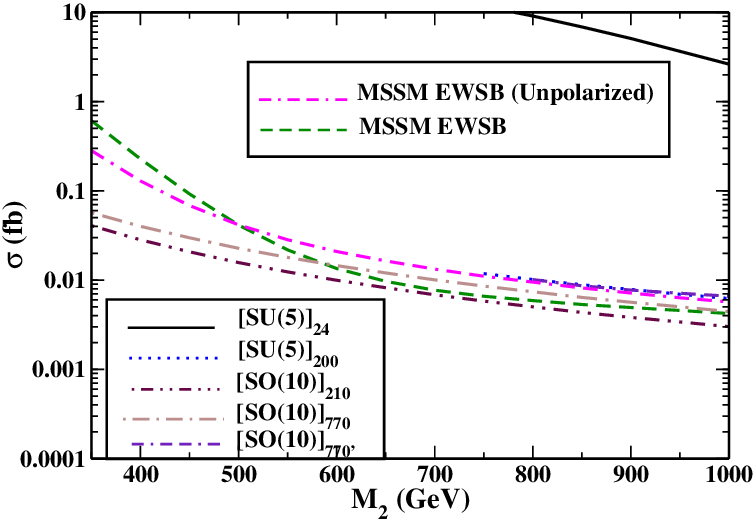}
\caption{Total cross section $\sigma$ with the inclusion of radiative effects
for the radiative neutralino production as a function of  $M_2$ for different
models with $M_2$ {\Large{$\epsilon$}} [390, 1000] GeV
at $\sqrt{s} = 500$~GeV and $(P_{e^-},P_{e^+})$ = (0.8, -0.6).
For comparison we have also shown the case of MSSM~EWSB with
unpolarized beams.}
\label{fig:M_2dep}
\end{minipage}
\end{figure}

In Fig.~\ref{fig:mu} we show the $\mu$ dependence of the cross section for
different models considered in this paper for the polarized case
along with the unpolarized case of MSSM EWSB. 
The cross section in the case of $SU(5)_{24}$ and MSSM EWSB is significantly
enhanced compared to the unpolarized case. 
For the other scenarios,
the behavior in case of polarized beams is almost similar to the unpolarized
case. It is found that for a wide range of 
 $\mu$, in the case of the $[SU(5)]_{24}$ and MSSM EWSB scenario,
all the  experimental constraints are satisfied, with $\tilde{\chi}_1^0$ as 
the LSP. For the other scenarios with a Higgsino-type lightest neutralino, the cross 
section is sensitive to the value of $\mu$. 
Since $m_{\tilde{\chi}_1^0} \propto \mu$, above  a certain value of 
$\mu$, $\tilde{\chi}_1^0$ ceases to be the lightest supersymmetric
particle. Depending on the percentage  of the Higgsino component, 
the cross section changes  with the value of $\mu$. 
Most of the scenarios considered here are tightly constrained  
as a function of $\mu$, with the neutralino as
the  LSP. This is due to the various limits  on the sparticles masses from the 
experiments. The cross section for some 
scenarios in this region  is too small to be observed  at the ILC with  
$\sqrt{s} = 500$~GeV, even with an integrated luminosity of $500$ fb$^{-1}$.


In Fig.~\ref{fig:M_2dep} we show the dependence of the radiative neutralino 
cross section
on the soft gaugino mass parameter $M_2$ for different models
with polarized beams. In this case also $SU(5)_{24}$ and MSSM EWSB show
an enhancement of the cross section, for smaller values of $M_2$. 
Since the total cross section decreases with increasing value of $M_2$,
a lower value of $M_2$ favors a cross section which can be measured 
experimentally.

\subsection{Dependence on selectron masses}

\begin{figure}[t]
\begin{minipage}[b]{0.45\linewidth}
\centering
\vspace*{0.7cm}
\includegraphics[width=6.5cm, height=5cm]{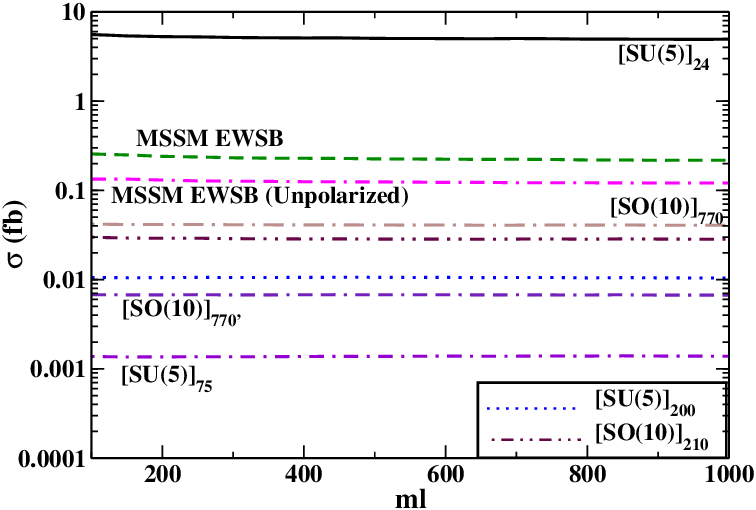}
\caption{Total cross section $\sigma$ for the radiative neutralino 
production with radiative effects included vs
$m_{\tilde{e}_L}$  at $\sqrt{s} = 500 $~GeV with $(P_{e^-},P_{e^+})$ 
= (0.8, -0.6). For comparison we have also shown the case of MSSM~EWSB 
with unpolarized beams.}
\label{fig:sneu_lt}
\end{minipage}
\hspace{0.4cm}
\begin{minipage}[b]{0.45\linewidth}
\centering
\includegraphics[width=6.5cm, height=5cm]{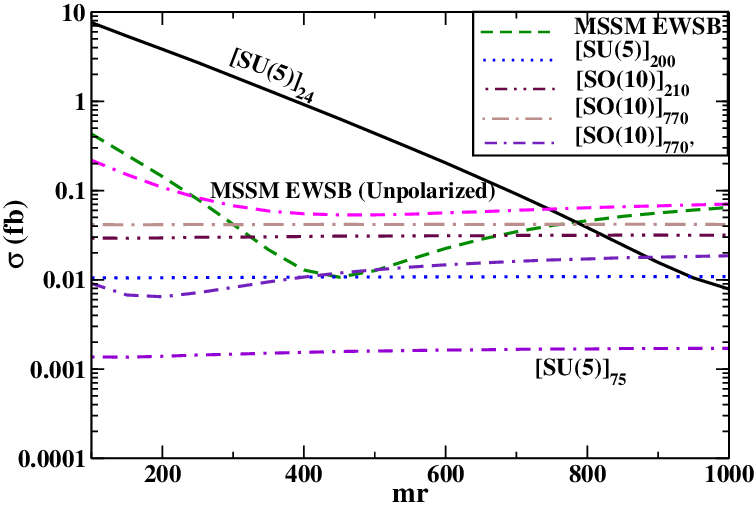}
\caption{Total cross section $\sigma$ along with radiative effects for 
the radiative neutralino production vs $m_{\tilde{e}_R}$  at 
$\sqrt{s} = 500 $~GeV with $(P_{e^-},P_{e^+})$ = (0.8, -0.6).
For comparison we have also shown the case of MSSM~EWSB with
unpolarized beams.}
\label{fig:sneu_rt}
\end{minipage}
\end{figure}

The selectron masses are free parameters for the models considered here.
Since the signal process proceeds mainly via right and left
selectron $\tilde e_{R,L}$ exchange in the $t$ and $u$ channels, we have also
considered the dependence of the total cross section on the selectron masses.
The dependence on the selectron masses is shown in Figs.~\ref{fig:sneu_lt} 
and  \ref{fig:sneu_rt} in the case of polarized beams and for unpolarized
beams in case of the MSSM EWSB. The cross section is insensitive to the left selectron
mass in case of all models. For $SU(5)_{24}$, the neutralino being a bino, the
cross section is sensitive to the right selectron mass and decreases with increasing
$m_{\tilde{e}_R}$ and has a negligible sensitivity to left selectron mass. 
The MSSM EWSB shows a peculiar behavior with respect to the right selectron mass.
This is mainly because the neutralino in this case has a dominant Higgsino component
along with a significant bino component. Therefore the signal process in this scenario
receives contribution from both the right selectron exchange channel and the
$Z$ exchange channel. This behavior arises due to the interference term from
these two diagrams and is sensitive to the centre-of-mass energy. Note that for
this particular choice of beam polarization, this behavior is more enhanced
as one of the contributing diagrams is due to $\tilde{e}_R$ exchange. If the beam
polarization would have been due to left-handed electrons and right-handed positrons,
there would be no contribution from the right selectron exchange diagram. Therefore
the cross section in that case will be insensitive to $m_{\tilde{e}_R}$.
The other models have a Higgsino-type neutralino; therefore their production 
cross section shows no dependence on the selectron masses. 

\section {Background Processes}
\label{sec:backgrounds}
\subsection{Neutrino background}\label{sec:neu}
\noindent

For the signal process~(\ref{radiative1}) considered here, the main
background comes from the SM radiative neutrino production. 
The other possible backgrounds are from $e^+e^- \rightarrow \tau^+ \tau^- \gamma$, with both the $\tau's$ decaying to soft leptons or hadrons
but the contribution from this process is found to be negligible. 
Another large background  comes  from the radiative Bhabha  scattering, 
$ e^+e^- \rightarrow  e^+ e^- \gamma$, where  ${e^\pm}'s$ are not
detected.  This  radiative scattering is usually eliminated by imposing 
a cut on $E_\gamma$. The 
events are selected by imposing the condition that any particle other 
than $\gamma$ appearing in the
angular range $-0.95 < \cos \theta_\gamma < 0.95$ must have energy less than 
$E_{max}$, where $E_{max}$ is detector dependent, but presumably
no larger than a few GeV. This is discussed in detail in 
the literature~\cite{Chen}.

\begin{figure}[htb]
\includegraphics{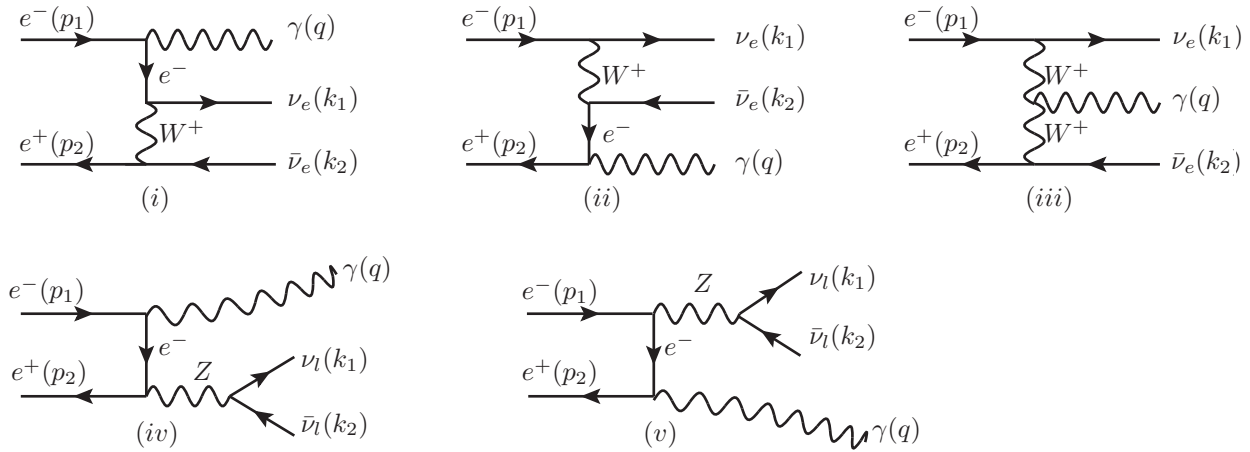}
\caption{Feynman diagrams contributing to the radiative neutrino process 
 $e^+e^- \rightarrow {\nu}{\bar{\nu}}\gamma$ where ($iv$ and $v$)
 corresponds to the neutrinos of three flavors}
\label{fig:radneutrino}
\end{figure}

The SM radiative neutrino production 
\begin{equation}
e^+ +e^- \to \nu_\ell+\bar\nu_\ell+\gamma\,,\;\;\qquad \ell=e,\mu,\tau
\label{radiative2}
\end{equation}
has been studied extensively~\cite{Datta:1996ur,Gaemers:1978fe,Berends:1987zz,
Boudjema:1996qg,Montagna:1998ce}.
For this background process, $\nu_e$ are produced via
$t$-channel $W$ boson exchange and $\nu_{e,\mu,\tau}$ via $s$-channel
$Z$ boson exchange.  The corresponding Feynman diagrams are shown in
Fig.~\ref{fig:radneutrino}.


\noindent

Since the photons emitted from this process mostly tend to be collinear,
therefore the angular cut on the photon is applied to separate it from the
signal photons. This process mainly proceeds through the exchange of $W$
bosons which couple only to the left-handed fermions. We are considering
the case of beam polarization with right-handed electron and left-handed positron.
The respective degree of polarization is $P_{e^-}$ = 0.8 and $P_{e^+}$ = -0.6.
Figure~\ref{fig:neu_pe} shows that the photon energy distribution from 
the radiative neutrino production, whereas in  Fig.~\ref{fig:neu_cs} we show
the $\sqrt s$ dependence of the total radiative neutrino cross section. Note that
the radiative corrections are included here. The unpolarized case is also shown
in the figures for comparison. It is observed that with this choice of
beam polarization, the $W$ bosons in the intermediate state do not contribute,
and the cross section is significantly reduced. For instance, at $\sqrt{s}$ = 500 GeV
with the inclusion of radiative corrections and the cuts, the total unpolarized
cross section $\sigma_{unpol}$ is 2432 fb, whereas with the inclusion of 
this particular beam polarization $\sigma_{pol}$ is 398 fb. The background is
reduced by 1 order of magnitude. Due to the production of $Z$ boson through the 
$s$ channel the photon energy distribution peaks for 
$E_\gamma= (s -4 m_{Z}^2)/(2\sqrt{s}) \approx 218$~GeV at $\sqrt{s}$ = 500 GeV.
By imposing an upper cut on the photon energy, which depends on the neutralino
mass see Eq.~(\ref{cut1}), the photon background
from radiative neutrino production is reduced.
A similar argument holds for the production cross section where the on-shell
$Z$ produced through this background process is eliminated by imposing an upper cut
on the photon energy.

\begin{figure}[htb]
\begin{minipage}[b]{0.4\linewidth}
\centering
\vspace*{0.7cm}
\includegraphics[width=6.5cm, height=5cm]{neutrino_photonenergy_n.eps}
\caption{Plot showing the photon energy distribution 
$d\sigma/dE_{\gamma}$
for the radiative neutrino production process
$e^+e^- \rightarrow \nu \bar\nu \gamma$ at $\sqrt{s} = 500$~GeV, with the 
inclusion of radiative effects and $(P_{e^-},P_{e^+})$ = (0.8, -0.6).}
\label{fig:neu_pe}
\end{minipage}
\hspace{0.7cm}
\begin{minipage}[b]{0.4\linewidth}
\centering
\includegraphics[width=6.5cm, height=5cm]{neutrino_cross_n.eps}
\caption{The total energy $\sqrt{s}$ dependence of the  radiative neutrino 
cross section with  and without an 
upper cut  on the photon energy $E_{\gamma}$, along with the
radiative effects and $(P_{e^-},P_{e^+})$ = (0.8, -0.6).}
\label{fig:neu_cs}
\end{minipage}
\end{figure}
\begin{figure}[htb]
{%
\unitlength=1.0pt
\includegraphics{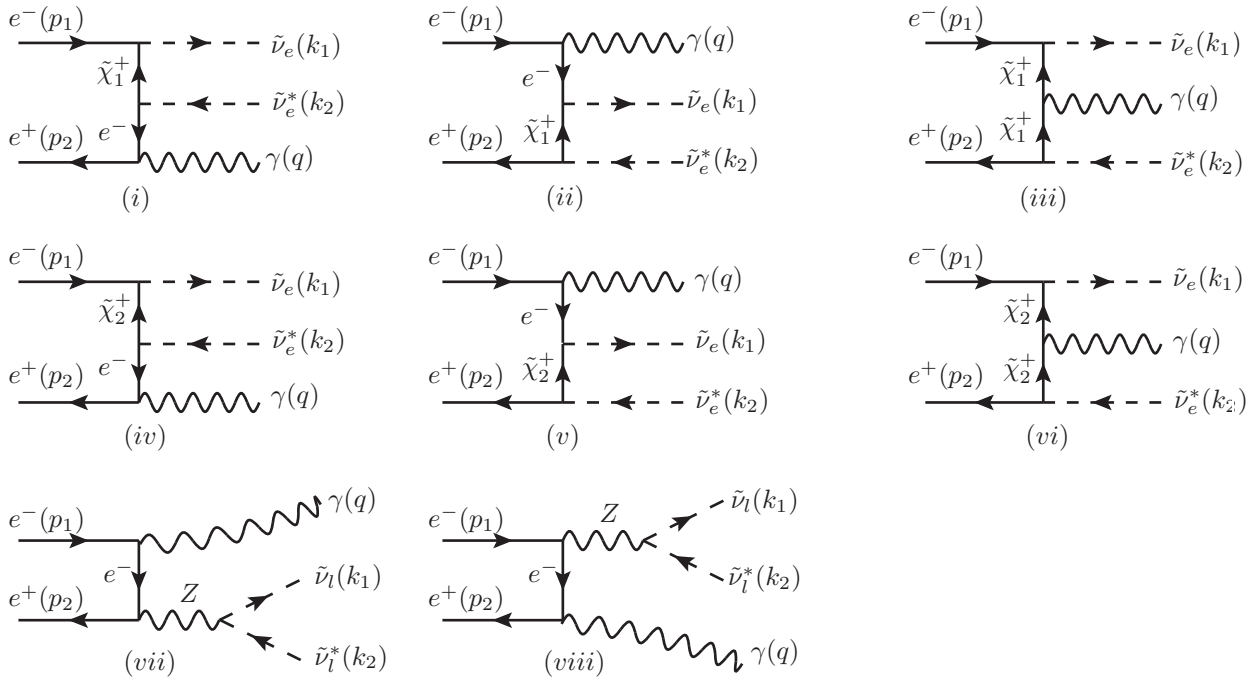}
}
\caption{Feynman diagrams contributing  to the radiative sneutrino production
process $e^+e^- \rightarrow \tilde{\nu}\tilde{\nu}^*\gamma$, with the last 
two diagrams ($vii$ and $viii$) corresponding to all the leptonic sneutrino }
\label{fig:radsneutrino}
\end{figure}
\noindent

\subsection{Supersymmetric background}

Apart from the SM background, the signal process  (\ref{radiative1}) under 
consideration has also a supersymmetric background from the sneutrino production 
process~\cite{Datta:1996ur, Franke:1994ph} 

\begin{equation}
e^+ +e^- \to \tilde\nu_\ell+\tilde\nu^\ast_\ell+\gamma\,, 
\;\qquad \ell=e,\mu,\tau\,.
\label{radiative3}
\end{equation}

\begin{figure}[htb]
\begin{minipage}[b]{0.4\linewidth}
\centering
\vspace*{0.7cm}
\includegraphics[width=6.5cm, height=5cm]{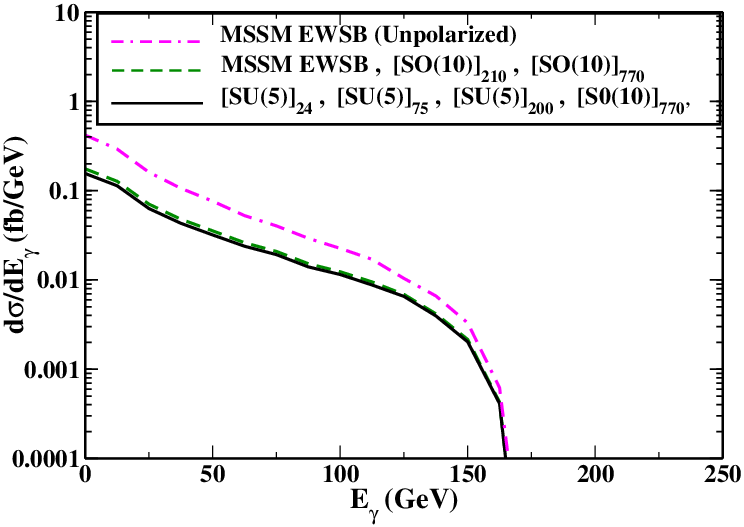}
\caption{Plot showing the photon energy distribution 
$d\sigma/dE_{\gamma}$
for the radiative sneutrino production process 
$e^+e^- \rightarrow \tilde{\nu}\tilde{\nu}^*\gamma$ at 
$\sqrt{s} = 500$~GeV, with the inclusion of radiative effects 
and initial beam polarization $(P_{e^-},P_{e^+})$ = (0.8, -0.6).
For comparison we have also shown the case of MSSM~EWSB with
unpolarized beams.}
\label{fig:sneu_pe}
\end{minipage}
\hspace{0.7cm}
\begin{minipage}[b]{0.4\linewidth}
\centering
\includegraphics[width=6.5cm, height=5cm]{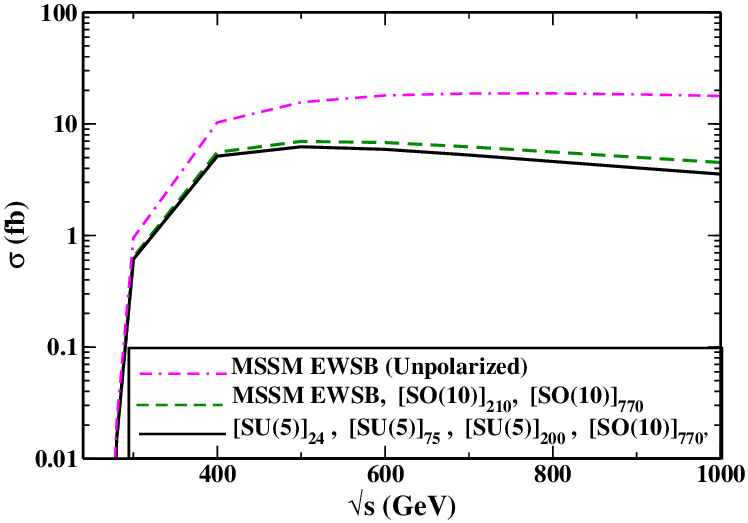}
\caption{The total energy $\sqrt{s}$  dependence of the radiative sneutrino 
cross section $e^+e^- \rightarrow \tilde{\nu}\tilde{\nu}^*\gamma$
with  an upper cut on the photon energy $E_\gamma$ and the inclusion 
of radiative effects and initial beam polarization 
$(P_{e^-},P_{e^+})$ = (0.8, -0.6). For comparison we have also 
shown the case of MSSM~EWSB with unpolarized beams.}
\label{fig:sneu_cs}
\end{minipage}
\end{figure}

\noindent In Fig.~\ref{fig:radsneutrino} we show the tree-level Feynman diagrams
contributing to the supersymmetric background process under study. 
Apart from the $s$  channel contribution from $Z$ boson, the process
also receives a $t$-channel contribution from the virtual charginos. Due to the
contribution from virtual charginos, this process is sensitive to the chargino mixing
matrix $\textbf{U}$. In Fig.~\ref{fig:sneu_pe} we show the photon energy
distribution for the supersymmetric background process at $\sqrt{s} = 500$~GeV for the
different models, whereas the total production cross section is shown in
Fig.~\ref{fig:sneu_cs}. We have applied the same cuts for this process as in the 
signal process and have used an initial
beam polarization of $P_{e^-}$ = 0.8, $P_{e^+}$ = -0.6.  
Similar to the radiative neutrino and neutralino production
the unpolarized case of MSSM EWSB is also included in the figures.
The process is not sensitive
to initial beam polarization, with the cross section and the photon energy
distribution in the case of polarized beams behaving 
almost similarly to the unpolarized case.
From  Figs.~\ref{fig:sneu_pe} and~\ref{fig:sneu_cs},  it is seen
that for $[SU(5)]_{24}, [SU(5)]_{75}, [SU(5)]_{200}$ and $[SO(10)]_{770'}$,
the behavior of the cross section is similar. This
is due to the mixing matrix $\bf{U}$ being same for all 
the models considered here. 

This process can act as a major supersymmetric background  to the 
signal if the sneutrinos decay invisibly via $\tilde\nu\to\tilde \chi^0_1\nu$.
This scenario has been called the ``virtual LSP'' scenario~\cite{Datta:1996ur}.
But the sneutrinos can decay to other particles if kinematically allowed 
thus reducing its contribution  to the signal. 
We note that
the other prominent decay channels are 
$\tilde\nu\to\tilde\chi^\pm_1\ell^\mp$ and $\tilde\nu\to\tilde \chi^0_2\nu$,
if kinematically allowed.
For the scenarios with a bino-type neutralino, the dominant decay mode is the 
invisible decay channel with 100\% branching ratio. For the scenarios with a 
Higgsino-type neutralino the various decay channels are presented in Table~\ref{br_neu}.

\begin{table}[htb]
\renewcommand{\arraystretch}{1.0}
\begin{center}
\vspace{0.5cm}
        \begin{tabular}{|c|c|c|c|c|c|c|}
\hline
Branching ratios &MSSM EWSB&$SU(5)_{75}$ &$SU(5)_{200}$ &$SO(10)_{210}$ &$SO(10)_{770}$ &$SO(10)_{770'}$\\ \hline \hline
${\rm  BR}(\tilde\nu_e\to\tilde\chi_1^0\nu_e)$ &78.4\%&8.1\% &21.2\%   &18\%   &24.2\%  &44.4\% \\ \hline
${\rm  BR}(\tilde\nu_e\to\tilde\chi_2^0\nu_e)$ &      &1.8\%  &4.54\%   &0.8\%   &1.2\%  &6\%   \\ \hline
${\rm  BR}(\tilde\nu\to\tilde\chi^\pm_1\ell^\mp)$&21.6\% &90.1\%  &74.3\%  &81\% &74.8\% &49.6\%   \\ \hline
\end{tabular}
\end{center}
\vspace{-0.5cm}
\caption{Branching ratios of the sneutrino for different models with a Higgsino-type lightest neutralino}
\renewcommand{\arraystretch}{1.0}
\label{br_neu}
\end{table}

There can also be other supersymmetric background from the neutralino production
$e^+e^- \to \tilde\chi_1^0\tilde\chi^0_2$, with the subsequent radiative 
decay~\cite{Haber:1988px} of
the next-to-lightest neutralino $\tilde\chi^0_2 \to \tilde\chi^0_1 \gamma$.
The branching ratios for this decay are too small, with a significant ratio 
obtained for small values of $\tan\beta<5$ or $M_1\sim M_2$ 
\cite{Ambrosanio:1995it,Ambrosanio:1995az,Ambrosanio:1996gz}. Therefore, we have neglected
this process in our study; however a detailed discussion of this process can be
found in  Refs.~\cite{Ambrosanio:1995az,Ambrosanio:1996gz,
Baer:2002kv}.


\subsection{Theoretical significance}

Finally we discuss  whether the photons from the signal
process can be measured over the photons from the background.  This is
expressed in terms of theoretical significance for a given integrated luminosity
$\mathcal{L}$ and is defined as~\cite{Dreiner:2006sb}
\begin{equation}
S  =  \frac{N_{\rm S}}{\sqrt{N_{\rm S} + N_{\rm B}}}=
\frac{\sigma}{\sqrt{\sigma + \sigma_{\rm B}}} \sqrt{\mathcal L}.
\label{significance}
\end{equation}
\noindent

\begin{figure}[htb]
\begin{minipage}[b]{0.4\linewidth}
\centering
\vspace*{0.7cm}
\includegraphics[width=6.5cm, height=5cm]{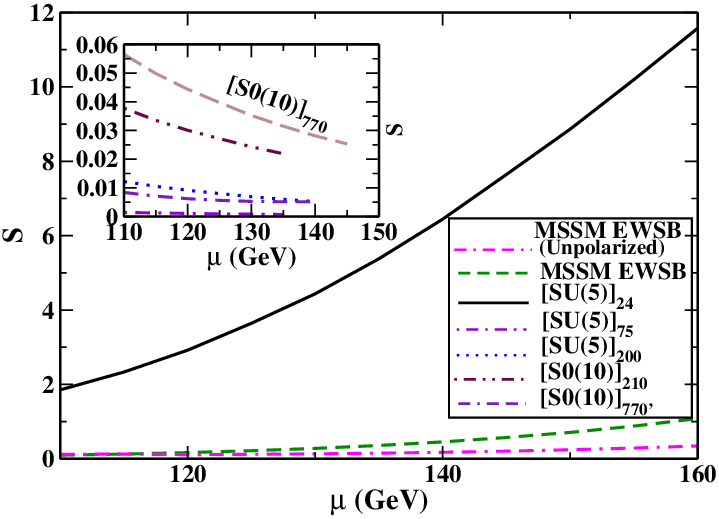}
\caption{Plot showing the theoretical significance 
$S$ for the radiative neutralino production as a function of  $\mu$ for 
different models considered in this paper with $\sqrt{s}= 500$~GeV
and $(P_{e^-},P_{e^+})$ = (0.8, -0.6). For comparison
we have also shown the case of MSSM~EWSB with unpolarized beams.}
\label{fig:sig_mu}
\end{minipage}
\hspace{0.7cm}
\begin{minipage}[b]{0.4\linewidth}
\centering
\includegraphics[width=6.5cm, height=5cm]{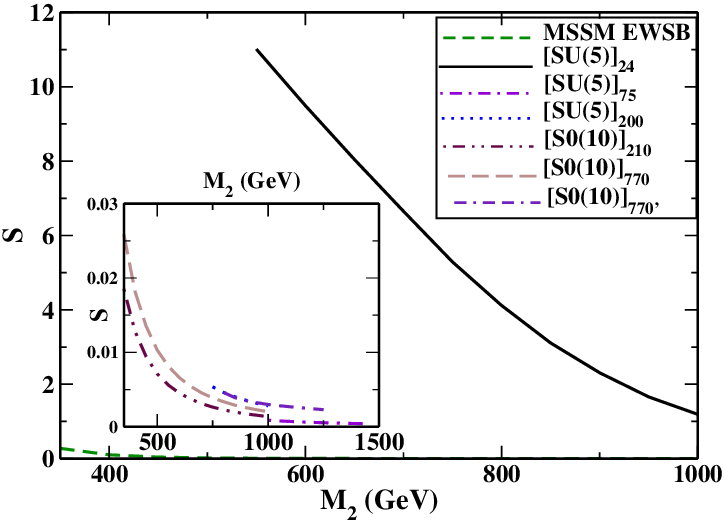}
\caption{The theoretical significance 
$S$ for the radiative neutralino production as a function of the gaugino 
mass parameter  $M_2$ for the different
models with  $\sqrt{s}= 500$~GeV and $(P_{e^-},P_{e^+})$ = (0.8, -0.6).
The case of unpolarized beams for MSSM~EWSB is not shown here as
it coincides with the polarized case.}
\label{fig:sig_M2}
\end{minipage}
\end{figure}

\noindent In the above equation $N_{\mathrm{S}}=\sigma {\mathcal L}$ is the
number of signal photons, and $N_{\rm B}=\sigma_{\rm B}{\mathcal L}$
denotes the number of background photons. For the detection of a signal 
a theoretical significance of 5 is required, whereas the signal can be
measured at a $68~\%$ confidence level for a theoretical significance of $S = 1$.
In Fig.~\ref{fig:sig_mu} we show the $\mu$ dependence of the theoretical
significance $S$ for the different models considered here for an initial
beam polarization of $P_{e^-}$ = 0.8 and $P_{e^+}$ = -0.6. When the
lightest neutralino is dominantly a bino  as in case of the $[SU(5)]_{24}$,
or has dominant bino and a Higgsino components,
as in the case of MSSM EWSB, the
choice of this beam polarization significantly enhances the signal compared to the
unpolarized case. In the case of unpolarized beams, $S$ for the considered $\mu$ range
has a maximum value of $2$ in the case of $[SU(5)]_{24},$ whereas for this choice of 
beam polarization, it has a maximum value of $12$. Similar behavior follows in case
of MSSM EWSB. It can be seen from the Fig.~\ref{fig:sig_mu} that it will be difficult
to observe the signal for the other scenarios considered here  with the lightest 
neutralino having a dominant Higgsino component.


We have also studied the variation of theoretical significance $S$ 
as a function of the gaugino mass parameter $M_2$. In Fig.~\ref{fig:sig_M2}
we show the $M_2$ dependence of $S$ for all the models considered in this work
in the interval $M_2$ {\Large{$\epsilon$}} [200,1000] GeV. A behavior almost
similar to the $\mu$ dependence of $S$ is observed. 


Along with $S$ we have also considered the signal-to-background ratio defined as
\begin{equation}
r=\frac{\sigma}{\sigma_B}
\label{r_ratio}
\end{equation}

\noindent The values of $S$ and $r$ can serve as a good guideline for our analysis
since we do not consider detector simulation here, which is beyond the scope
of the present paper. In the case of the ILC, for a signal to be detectable, $r$ is
required to be greater than 1\%. Since the future collider is designed for planned 
energies of 500, 800, and 1000 GeV, we have presented the signal and background 
cross sections along with $S$ and $r$ for these energies and different cases of a  
longitudinally polarized beam for an integrated luminosity of 
500 fb$^{-1}$. We present the values of the total cross section, the significance
and the signal-to-background ratio
for all the scenarios considered here for 
different center-of-mass energies in Tables~\ref{tab:S500},~\ref{tab:S800} and~\ref{tab:S1000}.
The set of parameters considered for the different models is listed in
Tables~\ref{parEWSB},~\ref{parEWSB24},~\ref{parEWSB75},~\ref{parEWSB200},~\ref{parSO(10)_210_1}
\ref{parSO(10)_770_1} and \ref{parSO(10)_770_2}. It can be seen from the 
Tables~\ref{tab:S500},~\ref{tab:S800} and~\ref{tab:S1000} that there is an 
enhancement in $S$ and $r$ when we move from the unpolarized to the polarized case.
The enhancement is significant for the case of beam polarization $(0.9|-0.6)$.
The behavior is similar with the bino-type neutralino having a significant
value of $S$ and $r$, making the signal observable at the ILC for different 
cases of beam polarization. But for the scenarios with a Higgsino-type
neutralino, the values of $S$ and $r$ are too small, making it difficult to test 
them at the future linear colliders through this radiative neutralino production process.

\begin{table}[htb]
\centering
\begin{tabular}{|c|c|c|c|c|c|c|c|c|}
\hline
 &$(P_{e^-}|P_{e^+})$ &$(0|0)$&$(0.8|0)$&$(0.8|-0.3)$
 &$(0.8|-0.6)$&$(0.9|0)$&$(0.9|-0.3)$&$(0.9|-0.6)$\\[1mm]
\hline
SM background &$\sigma(e^+e^-\rightarrow \nu \bar{\nu} \gamma)$ (fb) &2432 &577 &481 &398 &335
&314 &295 \\ [1mm] \hline
   &$\sigma(e^+e^-\rightarrow \tilde{\chi}_1^0 \tilde{\chi}_1^0 \gamma)$ (fb) &0.1377 &0.1651 &0.2096 &0.2495 &0.1704 &0.2172 &0.2601 \\ [1mm]
MSSM EWSB &S &0.0624 &0.1536 &0.2136 &0.2795 &0.2081 &0.2739 &0.3384    \\ [1mm]
  &r &0.0056 &0.0286 &0.0435 &0.0626 &0.0508 &0.0691 &0.0881     \\ [1mm] \hline
 &$\sigma(e^+e^-\rightarrow \tilde{\chi}_1^0 \tilde{\chi}_1^0 \gamma)$ (fb)          &1.883 &3.391  &4.432 &5.376 &3.551 &4.626 &5.758  \\ [1mm]
$[SU(5)]_{24}$    &S &0.8534 &3.1470 &4.4978 &5.9850 &4.3150 &5.7947 &7.4239     \\ [1mm]
&r &0.0774 &0.5876 &0.9214 &1.3507 &1.0600 &1.4732 &1.9518     \\ [1mm] \hline
  &$\sigma(e^+e^-\rightarrow \tilde{\chi}_1^0 \tilde{\chi}_1^0 \gamma)$ (fb)          &0.0007 &0.0009 &0.0011 &0.0014 &0.0009 &0.0012 &0.0014 \\ [1mm]
$[SU(5)]_{75}$&S &0.0003 &0.0008 &0.0011 &0.0015 &0.0010 &0.0015 &0.0018     \\ [1mm]
&r &0.0000 &0.0001 &0.0002 &0.0003 &0.0002 &0.0003 &0.0004     \\ [1mm] \hline
 &$\sigma(e^+e^-\rightarrow \tilde{\chi}_1^0 \tilde{\chi}_1^0 \gamma)$ (fb)          &0.0067 &0.0071 &0.0089 &0.0105 &0.0071 &0.0091 &0.0108 \\ [1mm]
$[SU(5)]_{200}$&S &0.0030 &0.0066 &0.0090 &0.0117 &0.0086 &0.0114 &0.0140     \\ [1mm]
&r &0.0002 &0.0012 &0.0018 &0.0026 &0.0021 &0.0028 &0.0036     \\ [1mm] \hline
 &$\sigma(e^+e^-\rightarrow \tilde{\chi}_1^0 \tilde{\chi}_1^0 \gamma)$ (fb) &0.0150 &0.0191 &0.0244 &0.0293 &0.0195 &0.0252 &0.0310 \\ [1mm]
$[SO(10)]_{210}$&S &0.0068 &0.0177 &0.0248 &0.0328 &0.0238 &0.0317 &0.0403     \\ [1mm]
&r &0.0006 &0.0033 &0.0050 &0.0073 &0.0058 &0.0080 &0.0105     \\ [1mm] \hline
 &$\sigma(e^+e^-\rightarrow \tilde{\chi}_1^0 \tilde{\chi}_1^0 \gamma)$ (fb) &0.0193 &0.0267 &0.0344 &0.0415 &0.0275 &0.0357 &0.0440 \\ [1mm]
$[SO(10)]_{770}$&S &0.0087 &0.0248 &0.0350 &0.0465 &0.0335 &0.0450 &0.0572     \\ [1mm]
&r &0.0007 &0.0047 &0.0071 &0.0104 &0.0082 &0.0136 &0.0149     \\ [1mm] \hline
 &$\sigma(e^+e^-\rightarrow \tilde{\chi}_1^0 \tilde{\chi}_1^0 \gamma)$ (fb) &0.0117 &0.0055 &0.0061 &0.0066 &0.0048 &0.0056 &0.0065 \\ [1mm]
$[SO(10)]_{770'}$&S &0.0053 &0.0051 &0.0062 &0.0073 &0.0058 &0.0070 &0.0084     \\ [1mm]
&r &0.0004 &0.0009 &0.0012 &0.0016 &0.0014 &0.0017 &0.0022     \\ [1mm] \hline 
\end{tabular}
\caption{ Signal and background cross sections $\sigma$, significance $S$,
 and signal-to-background ratio $r$ in the case of different beam polarizations
 $(P_{e^-}| P_{e^+})$ for the different scenarios  at $\sqrt{s}$ = 500 GeV
 for $\mathcal{L}$ = 500 fb$^{-1}$.}
\label{tab:S500}
\end{table} 

\begin{table}[htb]
\centering
\begin{tabular}{|c|c|c|c|c|c|c|c|c|}
\hline
 &$(P_{e^-}|P_{e^+})$ &$(0|0)$&$(0.8|0)$&$(0.8|-0.3)$
 &$(0.8|-0.6)$&$(0.9|0)$&$(0.9|-0.3)$&$(0.9|-0.6)$\\[1mm]
\hline
SM background &$\sigma(e^+e^-\rightarrow \nu \bar{\nu} \gamma)$ (fb) & 2365 &505 &375 &284 &280
 &230 &177\\ [1mm] \hline
 &$\sigma(e^+e^-\rightarrow \tilde{\chi}_1^0 \tilde{\chi}_1^0 \gamma)$ (fb) & 0.1002 &0.1343 &0.1690 &0.2055 &0.1371 &0.1772 &0.2168 \\ [1mm]
MSSM EWSB&S &0.0460 &0.1336 &0.1950 &0.2725 &0.1831 &0.2611 &0.3641     \\ [1mm]
&r &0.0042 &0.0265 &0.0450 &0.0723 &0.0489 &0.0770 &0.1224     \\ [1mm] \hline 
 &$\sigma(e^+e^-\rightarrow \tilde{\chi}_1^0 \tilde{\chi}_1^0 \gamma)$ (fb)          & 1.4103 &2.5440 &3.3289 &4.0442 &2.6612 &3.4729 &4.3246 \\ [1mm]
$[SU(5)]_{24}$&S &0.6482 &2.5249 &3.8268 &5.3280 &3.5392 &5.0821 &7.1810     \\ [1mm]
&r &0.0596 &0.5037 &0.8877 &1.4240 &0.9504 &1.5100 &2.4433    \\ [1mm] \hline 
&$\sigma(e^+e^-\rightarrow \tilde{\chi}_1^0 \tilde{\chi}_1^0 \gamma)$ (fb)          &0.0004 & 0.0005 &0.0006 &0.0008 &0.0005 &0.0007 &0.0008 \\ [1mm]
$[SU(5)]_{75}$ &S &0.0001 &0.0004 &0.0006 &0.0010 &0.0006 &0.0010 &0.0013     \\ [1mm]
&r &0.0000 &0.0000 &0.0001 &0.0002 &0.0001 &0.0003 &0.0004     \\ [1mm] \hline 
 &$\sigma(e^+e^-\rightarrow \tilde{\chi}_1^0 \tilde{\chi}_1^0 \gamma)$ (fb)          &0.0037 &0.0040 & 0.0051 &0.0060 &0.0042 &0.0052 &0.0063 \\ [1mm]
$[SU(5)]_{200}$&S &0.0017 &0.0039 &0.0058 &0.0079 &0.0056 &0.0076 &0.0105     \\ [1mm]
&r &0.0001 &0.0007 &0.0013 &0.0021 &0.0015 &0.0022 &0.0035     \\ [1mm] \hline 
 &$\sigma(e^+e^-\rightarrow \tilde{\chi}_1^0 \tilde{\chi}_1^0 \gamma)$ (fb) &0.0097 & 0.0111 &0.0140 &0.0167 &0.0112 &0.0144 &0.0176 \\ [1mm]
$[SO(10)]_{210}$&S &0.0044 &0.0110 &0.0161 &0.0221 &0.0149 &0.0212 &0.0295     \\ [1mm]
&r &0.0004 &0.0021 &0.0037 &0.0058 &0.0040 &0.0062 &0.0099     \\ [1mm] \hline 
&$\sigma(e^+e^-\rightarrow \tilde{\chi}_1^0 \tilde{\chi}_1^0 \gamma)$ (fb) &0.0112 &0.0154 &0.0197 &0.0238 &0.0158 &0.0205 &0.0252 \\ [1mm]
$[SO(10)]_{770}$ &S &0.0051 &0.1532 &0.0227 &0.0315 &0.0211 &0.0302 &0.0423     \\ [1mm]
&r &0.0004 &0.0304 &0.0052 &0.0083 &0.0056 &0.0089 &0.0142     \\ [1mm] \hline 
 &$\sigma(e^+e^-\rightarrow \tilde{\chi}_1^0 \tilde{\chi}_1^0 \gamma)$ (fb) &0.0067 &0.0032 &0.0035 &0.0038 &0.0027 &0.0033 &0.0038\\ [1mm]
$[SO(10)]_{770'}$&S &0.0030 &0.0031 &0.0040 &0.0050 &0.0036 &0.0048 &0.0063     \\ [1mm]
&r &0.0002 &0.0006 &0.0009 &0.0013 &0.0009 &0.0014 &0.0021     \\ [1mm] \hline 

\end{tabular}
\caption{ Signal and background cross sections $\sigma$, significance $S$,
 and signal-to-background ratio $r$ in the case of different beam polarizations
 $(P_{e^-}| P_{e^+})$ for the different scenarios  at $\sqrt{s}$ = 800 GeV
 for $\mathcal{L}$ = 500 fb$^{-1}$.}
\label{tab:S800} 
\end{table} 
\begin{table}[htb]
\centering
\begin{tabular}{|c|c|c|c|c|c|c|c|c|}
\hline
 &$(P_{e^-}|P_{e^+})$ &$(0|0)$&$(0.8|0)$&$(0.8|-0.3)$
 &$(0.8|-0.6)$&$(0.9|0)$&$(0.9|-0.3)$&$(0.9|-0.6)$\\[1mm]
\hline
 SM background &$\sigma(e^+e^-\rightarrow \nu \bar{\nu} \gamma)$ (fb) &2293 &482 &356 &248 &257
 &204 &155\\ [1mm] \hline
  &$\sigma(e^+e^-\rightarrow \tilde{\chi}_1^0 \tilde{\chi}_1^0 \gamma)$ (fb) 
  & 0.0781 &0.1073 &0.1378 &0.1652 &0.1100 &0.1431 &0.1759 \\ [1mm]
MSSM EWSB &S &0.0364 &0.1092 &5.8147 &0.2344 &0.1533 &0.2239 &0.3157     \\ [1mm]
&r &0.0034 &0.0222 &0.0387 &0.0666 &0.0428 &0.0701 &0.1134     \\ [1mm] \hline 
 &$\sigma(e^+e^-\rightarrow \tilde{\chi}_1^0 \tilde{\chi}_1^0 \gamma)$ (fb)          &1.1022 &1.9900 &2.6008 &3.1510 & 2.0845 &2.7096 &3.3768 \\ [1mm]
$[SU(5)]_{24}$&S &0.5145 &2.0225 &3.0709 &4.4338 &8.6425 &4.2140 &5.999     \\ [1mm]
&r &0.0480 &0.4128 &0.7305 &1.2705 &0.8110 &1.328 &2.1786     \\ [1mm] \hline 
 &$\sigma(e^+e^-\rightarrow \tilde{\chi}_1^0 \tilde{\chi}_1^0 \gamma)$ (fb)          &0.0003 & 0.0004 &0.0005 &0.0005 &0.0004 &0.0005 &0.0006 \\ [1mm]
$[SU(5)]_{75}$&S &0.0001 &0.0004 &0.0006 &0.0007 &0.0005 &0.0007 &0.0010     \\ [1mm]
&r &0.0000 &0.0000 &0.0001 &0.0002 &0.0002 &0.0003 &0.0004     \\ [1mm] \hline 
 &$\sigma(e^+e^-\rightarrow \tilde{\chi}_1^0 \tilde{\chi}_1^0 \gamma)$ (fb)          &0.0027 & 0.0029 &0.0036 &0.0043 &0.0029 &0.0037 &0.0045 \\ [1mm]
$[SU(5)]_{200}$&S &0.0012 &0.0029 &0.0042 &0.0061 &0.0040 &0.0057 &0.0080     \\ [1mm]
&r &0.0001 &0.0006 &0.0010 &0.0017 &0.0011 &0.0018 &0.0029     \\ [1mm] \hline 
&$\sigma(e^+e^-\rightarrow \tilde{\chi}_1^0 \tilde{\chi}_1^0 \gamma)$ (fb) &0.0074 &0.0081 &0.0100 & 0.0121 &0.0081 &0.0104 &0.0126 \\ [1mm]
$[SO(10)]_{210}$ &S &0.0034 &0.0082 &0.0118 &0.1561 &0.0112 &0.0162 &0.0226     \\ [1mm]
&r &0.0003 &0.0017 &0.0028 &0.0017 &0.0032 &0.0051 &0.0081     \\ [1mm] \hline 
 &$\sigma(e^+e^-\rightarrow \tilde{\chi}_1^0 \tilde{\chi}_1^0 \gamma)$ (fb) &0.0082 &0.0111 &0.0142 &0.0171 &0.0114 &0.0147 &0.0181 \\ [1mm]
$[SO(10)]_{770}$&S &0.0038 &0.0113 &0.0168 &0.0242 &0.0159 &0.0229 &0.0325     \\ [1mm]
&r &0.0004 &0.0023 &0.0040 &0.0069 &0.0044 &0.0072 &0.0116     \\ [1mm] \hline 
 &$\sigma(e^+e^-\rightarrow \tilde{\chi}_1^0 \tilde{\chi}_1^0 \gamma)$ (fb) &0.0048 &0.0023 &0.0026 &0.0028 &0.002 &0.0024 &0.0028 \\ [1mm]
$[SO(10)]_{770'}$&S &0.0048 &0.0023 &0.0030 &0.0039 &0.0027 &0.0037 &0.0050     \\ [1mm]
&r &0.0002 &0.0005 &0.0007 &0.0011 &0.0008 &0.0012 &0.0032     \\ [1mm] \hline 
\end{tabular}
 \caption{Signal and background cross sections $\sigma$, significance $S$,
 and the signal-to-background ratio $r$ in the case of different beam polarizations
 $(P_{e^-}| P_{e^+})$ for the different scenarios  at $\sqrt{s}$ = 1000 GeV
 for $\mathcal{L}$ = 500 fb$^{-1}$.}
 \label{tab:S1000}
\end{table} 


\subsection{Left-right asymmetry}
\label{sec:lrasymmetry}
\begin{figure}[htb]
\centering
\vspace{0.6cm}
\includegraphics[width=6.5cm, height=5cm]{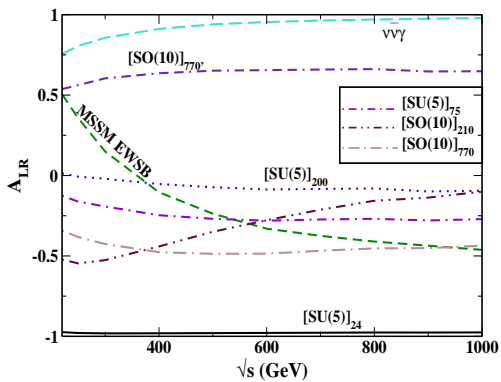}
\caption{Plot showing the left-right asymmetry as a function of the 
center-of-mass energy for radiative neutralino production
in the case of different models along with the SM background from radiative 
neutrino production.}
\label{fig:lrasym}
\end{figure}

In this subsection, we consider as an observable 
the left-right asymmetry as a means to distinguish between various 
grand unified models. In order to obtain a better efficiency,
we consider the integral version of this asymmetry.
The integrated
left-right asymmetry is defined as
\begin{equation}
A_{LR}= \frac{\sigma_{LR}-\sigma_{RL}}{\sigma_{LR}+\sigma_{RL}},
\end{equation}
where $\sigma_{RL}$ and $\sigma_{LR}$ are defined in
Sec.~\ref{subsec:LP}. The coupling of the lightest neutralino to
the selectron and a 
fermion is different for left- and right- handed fermions, with
different couplings, for the different models that we have considered
in this paper. The coupling is relatively sensitive to the
composition of the lightest neutralino, and one would expect an
appreciable difference between the left- and right-polarized
cross section. It can be seen from Table~\ref{feynmandiag}
that the lightest neutralino with a dominant wino and bino composition
is sensitive to beam polarization, whereas the Higgsino type neutralino
has no dependence on beam polarization.
We plot in Fig.~\ref{fig:lrasym} the left-right asymmetry for the
different models for the  radiative neutralino production
as a function of the centre-of-mass energy.
The SM background (radiative neutrino production) is also considered
here. The dependence of the various models on beam polarization can be
easily understood from Fig.~\ref{fig:lrasym}. In the case of radiative
neutrino production, as discussed before, since the cross section gets
highly suppressed with positive electron and negative positron beam
polarization, $A_{LR}$ in this case is the largest. $A_{LR}$ is also
larger for the models where the lightest
neutralino is mainly a bino or a wino, i.e., enhanced couplings to the
selectrons. Since $SU(5)_{24},~S0(10)_{770'}$ have a bino- and wino-type
lightest neutralino, $A_{LR}$, in this case is close to 1 or greater than
0.5. But for most of the models ($SU(5)_{75},~SU(5)_{200},~SO(10)_{210},
S0(10)_{770},~S0(10)_{770'}$), $A_{LR}$ is less than 0.5, since
they have the lightest neutralino with a dominant
Higgsino component, with practically no beam polarization dependence
from the exchange of selectrons. The result is almost independent of the
center-of-mass energy.

\section{Summary and Conclusions}
\label{sec:conclusions}
In this paper we have carried out a detailed study of the radiative neutralino
production $e^+e^- \to\tilde\chi^0_1 \tilde\chi^0_1\gamma$
for the case of $SU(5)$ and $SO(10)$ supersymmetric GUT models for ILC
energies with longitudinally polarized $e^-$ and $e^+$ beams.
In these GUT models the boundary conditions
on the soft gaugino mass parameters can be nonuniversal. 
We have compared the results
of these GUT models with the corresponding results in the MSSM with
universal gaugino mass parameters~(universal boundary conditions).
For our analyses we have used
a particular set of parameter values for various models by imposing
theoretical and experimental constraints as discussed 
in Sec.~\ref{sec:exp_cons}.
The radiative neutralino production process has a signature of a 
high-energy photon and missing energy. The background to the signal 
process comes from the SM process
$e^+e^-\rightarrow \nu \bar{\nu}\gamma$ and from the supersymmetric
process $e^+e^-\rightarrow \tilde{\nu}\tilde{\nu}^*\gamma$. 


The purpose of the present work is to establish  the
use of longitudinal beam polarization in probing the 
effects of boundary conditions in the neutralino sector
that arise in GUTs at a linear collider.  
This is motivated by the fact that longitudinal polarization is a distinct
possibility at the ILC. For the signal process considered, the dominant 
SM background comes from the  radiative neutrino production process,
which proceeds through 
the exchange of $W$ bosons which couple only to the left-handed particles. 
This dominant background made 
it difficult  to observe the signal process at  the LEP even for very light 
neutralinos.  At the LHC also the CMS experiment has searched for a final state 
containing a photon and missing transverse energy, and the observed
event yield was seen to be in agreement with the standard model
expectations.
However in the case of the  ILC with the availability of beam
polarization, a suitable choice of beam polarization ($e^-_R e^+_L$) 
will significantly reduce the expected SM  background. 
Therefore, the ILC, with the availability of beam polarization,
will be a good place to look for the processes with
a high-energy  photon and large missing energy in the final state.
At the future linear colliders, because of high luminosity, ISR 
and beamstrahlung are an unavoidable feature, and, therefore, 
we have included the radiative corrections in our calculations
to obtain a precise values for the cross sections.

We have studied in detail the cross section and the photon energy
distribution for the signal and background process for a centre-of-mass
energy of 500 GeV and an integrated luminosity of 500 fb$^{-1}$. 
The initial beams are taken to be longitudinally polarized
with $P_{e^-}$ = 0.8 and $P_{e^+}$ = -0.6.
Our analyses  show the behavior of the different models with the 
inclusion of beam polarization.  Together with these
the dependence of the cross section on the other free SUSY parameters 
which are involved in the signal process was also studied. 
This includes  the $SU(2)_L$ gaugino mass parameter $M_2$ and 
the Higgs(ino) mass parameter $\mu$ as well as the selectron
masses ($m_{\tilde{e}_R},m_{\tilde{e}_L}$).
Our results demonstrate that  the composition of the
lightest neutralino in different models plays a crucial role 
in the signal process.  It can be seen from Table~\ref{feynmandiag}
how the bino- and wino-type neutralino production cross section
will be controlled by  different choices of initial beam polarization. 
Similarly the insensitivity of the Higgsino-type neutralino production
cross section to the beam polarization, which is mediated through thr $Z$ boson, 
is also reflected in the table.
For the bino-type neutralino which arises in $SU(5)_{24}$, with 
significantly larger coupling to the right selectron, the cross section is 
increased with the choice of beam polarization used here, and the 
background is correspondingly reduced. At the same time in the 
case of other models, with Higgsino-type lightest neutralino,
there is no appreciable change in cross section for the  choice of beam
polarization used in this paper, since  $Z$ couples to both left- and 
right-handed fermions. 

Finally, in order
to study whether an excess of signal photons $N_S$ can be observed 
over the background photons $N_B$ from the SM radiative neutrino process, 
we have studied the theoretical statistical significance $S$ and the 
signal-to-background ratio $r$. The dependence
of $S$ on the independent parameters $M_2$ and $\mu$ is also studied.
The results that we have obtained  emphasize the signal
and the background cross sections along with the significance and the 
signal-to-background ratio for  different degrees of initial beam polarization
at different planned centre-of-mass energies of the ILC. They are presented
in Tables~\ref{tab:S500},~\ref{tab:S800}, and ~\ref{tab:S1000}.
Therefore, we conclude that in the presence of beam polarization with right-handed 
electrons and left-handed positrons,  the models with a bino-type 
neutralino can be studied in detail through the radiative neutralino 
production it the ILC. 
In this respect the grand unified supersymmetric  $SU(5)_{24}$ model is unique among all 
the  models considered in this paper. In this case $M_3$ can be 
large so that 
we get the gluino mass satisfying the experimental constraints, and also
$M_1$ will be small  enough to lead to a light bino-type neutralino. 
Therefore, for the choice of parameters considered 
in our paper, $SU(5)_{24}$ will provide a signal
which could be  observed at the ILC. This provides a strong motivation
for the search for the radiative neutralino production as an evidence
of a supersymmetric grand unified model at the ILC.

We would also like to point that even with initially polarized beams,
the models with a Higgsino-type neutralino will be too difficult to 
be observed at the ILC.
These Higgsino-type scenarios have a distinctive feature
wherein $\tilde{\chi}_1^0, \tilde{\chi}_2^0$ and $\tilde{\chi}_1^\pm$ are
almost degenerate
with masses around $\mu$ due to large values of $M_{1,2}$ and a low value
of $\mu$.
Due to the degeneracy in mass, the processes
(a) $e^+e^- \rightarrow {\tilde {\chi}_1^0}{\tilde {\chi}_2^0}\gamma$ and
(b) $e^+e^- \rightarrow {\tilde {\chi}_2^0}{\tilde {\chi}_2^0}\gamma$
will also yield a similar final state as the radiative neutralino production.
A detailed study of signatures with a hard photon and large missing energy
will include processes a and b along with the signal process considered here.
This will result in a significant increase of cross section, and may offer
additional search avenues. We note here that
${\tilde {\chi}_1^0}{\tilde {\chi}_2^0}$ and
${\tilde {\chi}_2^0}{\tilde {\chi}_2^0}$ production channels tend to be
suppressed, but may, nevertheless, offer increased search avenues.
We do not consider this case any further here,
but leave it for a future publication.

\section{acknowledgements}
The authors would like to thank B. Ananthanarayan for many
useful discussions.  
P.~N.~P. would like to thank the Centre for High Energy Physics, 
Indian Institute of Science, Bangalore for  hospitality while this
work was initiated. The work of P.~N.~P. is supported by the 
J. C. Bose National Fellowship of the Department of Science and Technology,
and by the Council of Scientific and Industrial Research,
India, under project No.~(03)(1220)/12/EMR-II. 
P.~N.~P would like to thank the Inter-University Centre 
for Astronomy and Astrophysics, Pune, India for hospitality
where part of this work carried out.
\appendix

\section{ GAUGINO MASSES  IN GRAND UNIFIED THEORIES}
\label{sec:gaugino mass patterns}
In this section we review the 
nonuniversal and universal gaugino masses
in grand unified theories. 

\subsection{Universal gaugino masses in  grand unified theories}
\label{subsec:universal-gaugino-masses} 

In  supersymmetric models, with gravity 
mediated supersymmetry breaking,
usually denoted as mSUGRA, the soft supersymmetry 
breaking gaugino mass parameters 
$M_1, M_2$, and $M_3$  
satisfy the universal boundary
conditions 
\bea
M_1 & = &  M_2 = M_3 = m_{1/2} \label{gauginogut}
\eea
at the grand unified scale $M_G$, where $ i = 1, 2, 3 $ refer to the 
$U(1)_Y, SU(2)_L$, and the $SU(3)_C$ gauge groups, respectively. 
Furthermore, the three gauge couplings corresponding to these gauge 
groups satisfy~($\alpha_i = g_i^2/4\pi, \, i = 1, 2, 3$) 
\bea
\alpha_1 & = & \alpha_2 =  \alpha_3 = \alpha_G,  \label{gaugegut}
\eea
at the GUT scale $M_G$,
where $g_1 =\frac{5}{3}g',\; g_2 = g$, with $g'$ and $g$ as 
$U(1)_Y$, and $SU(2)_L$ gauge couplings, respectively, and $g_3$ 
is the $SU(3)_C$ gauge coupling.  
The renormalization group equations then imply that imply that out of 
three gaugino mass parameters,
only one is independent, which we are free to choose as the gluino 
mass $M_3 \equiv M_{\tilde g}$.   
For the gaugino mass parameters, this leads to the ratio
\begin{equation}
M_1 : M_2 : M_3 \simeq 1 : 2 : 7.1.
\label{msugra0}
\end{equation}
The gaugino mass parameters described above are the running masses
evaluated at the electroweak scale $M_Z$. A lower bound is then obtained on
the parameter $M_1$ in the case of mSUGRA, from the constraint on $M_2$ (\ref{limits1})
and the ratio (\ref{msugra0}) :
\bea
M_1 & \gsim & 50~ {\rm GeV} \label{msugra2}
\eea

\subsection{Nonuniversal gaugino masses in  grand unified theories}
\label{subsec:non-universal-gaugino-masses}

In contrast to the Sec.~\ref{subsec:universal-gaugino-masses} with universal boundary 
condition~(\ref{gauginogut}) for the gaugino mass parameters at the GUT scale,
we now consider the case of MSSM with nonuniversal boundary conditions 
at the GUT scale, which arise in $SU(5)$ and $SO(10)$ grand unified theories.
Since in supersymmetric GUTs the gaugino masses need not  be 
equal at the GUT scale, the neutralino masses and mixing can be different in SUSY
GUTs as compared to the MSSM with universal gaugino masses.

The coupling of the field strength superfield $W^a$ with the gauge kinetic function
$f(\Phi)$ results in the generation of soft gaugino masses in 
supersymmetric models (see Ref.~\cite{Pandita:2012es} and references therein). 
This term can be written as
\bea
{\cal L}_{g.k.} \; & = & \;
\int d^2\theta f_{ab}(\Phi) W^{a}W^{b}
+h.c.,
\label{gk}
\eea
with $a$ and $b$ referring to gauge group indices and repeated indices being
summed over. The gauge kinetic function $f_{ab}(\Phi)$ can be written in
terms of the singlet and nonsinglet chiral superfields. 

When the auxiliary part $F_\Phi$ of a chiral superfield
$\Phi$ in $f(\Phi)$ gets a VEV $\langle F_\Phi \rangle$,
the interaction~(\ref{gk}) gives rise to soft gaugino masses:
\bea
{\cal L}_{g.k.} \; \supset \;
{{{\langle F_\Phi \rangle}_{ab}} \over {M_P}}
\lambda^a \lambda^b +h.c.,
\eea
where $\lambda^{a,b}$ are gaugino fields. Here 
$\lambda^1$, $\lambda^2$, and $\lambda^3$ are  the
$U(1)$, $SU(2)$, and $SU(3)$
gaugino fields, respectively. Since the gauginos belong to the adjoint
representation of the gauge group,
$\Phi$ and $F_\Phi$ can belong to any of the
representations appearing in the symmetric product of the
two adjoint  representations of the corresponding gauge group.
We note that in four-dimensional grand unified theories only
the gauge groups $SU(5)$, $SO(10)$, and  $E_6$ support the chiral
structure of weak interactions. Here we shall study the
implications of nonuniversal gaugino masses  for the case of
$SU(5)$ and  $SO(10)$ grand 
unified gauge groups.

\subsubsection{$SU(5)$}
\label{non-universal-SU(5)}
In this section we shall consider the case where the  SM gauge group 
is embedded in the  grand unified gauge group $SU(5)$.
For the symmetric product of the two adjoint~({\bf 24}-dimensional)
representations of $SU(5)$, we have
\bea
({\bf 24 \otimes 24})_{Symm} = {\bf 1 \oplus 24 \oplus 75 \oplus 200}.
\label{product}
\eea
In the simplest case where $\Phi$ and $F_\Phi$ are
assumed to be in the singlet representation of $SU(5)$, we have
equal gaugino masses at the GUT scale.  But, as is obvious from 
Eq.~(\ref{product}), $\Phi$ and  $F_\Phi$
can belong to any of the nonsinglet representations
{\bf 24}, {\bf 75},  and {\bf 200} of $SU(5)$. In such  cases 
the soft gaugino masses are unequal but related to one another via the
representation invariants of the gauge group~\cite{Ellis:1985jn}.  
In Table~\ref{tab1} we show the ratios of  gaugino masses 
which result when $F_{\Phi}$ belong to different representations of 
$SU(5)$ in the decomposition~(\ref{product}).
In this paper, for definiteness, we shall study the case of each 
representation independently, although an arbitrary combination of these is 
also allowed.

In the one-loop approximation, the solution of renormalization 
group equations for the soft supersymmetry breaking
gaugino masses $M_1$, $M_2$, and $M_3$  can be written 
as~\cite{Martin:1993ft}
\bea {{M_i (t)} \over {\alpha_i(t)}} = {{{M_i}({\rm GUT})} \over
{{\alpha_i}({\rm GUT})}}, \, \, \, i = 1, 2, 3. 
\label{rel1} \eea
\noindent Then at  any arbitrary scale, we have
\bea
{M_1} = {\frac 5 3}{{\alpha} \over {\cos^2{\theta_W}}}
\left({{{M_1}({\rm GUT})} \over {{\alpha_1}({\rm GUT})}}\right),\;\;
{M_2} = {{\alpha} \over {\sin^2{\theta_W}}}
\left({{{M_2}({\rm GUT})} \over {{\alpha_2}({\rm GUT})}}\right),\;\;
{M_3} = {\alpha_3} \left({{{M_3}({\rm GUT})} \over
{{\alpha_3}({\rm GUT})}}\right).\nonumber\\
\label{rel2}
\eea


\begin{table}[t!]
\renewcommand{\arraystretch}{1.0}
\begin{center}
  \begin{tabular}{||c|ccc|ccc||}
    \hline 
    $SU(5)$ & $M_1^G$ & $M_2^G$ & $M_3^G$ & 
    $M_1^{EW}$ & $M_2^{EW}$ & $M_3^{EW}$
    \\ \hline 
    {\bf 1} & 1 & 1
    & 1 & 1 & 2 & 7.1 \\ 
    & & & & & & \\    
    {\bf 24} & 1 & 3 & -2 & 1 & 6 & -14.3 \\
     & & & & & & \\    
     {\bf 75} & 1 &-$\frac{3}{5}$ &-$\frac{1}{5}$ & 1 & -1.18 & -1.41 \\
      & & & & & & \\    
      {\bf 200} & 1 & $\frac{1}{5}$ &$\frac{1}{10}$ &1 & 0.4 & 0.71
    \\ \hline
  \end{tabular}
  \end{center}
  \caption{\label{tab1} Ratios of the gaugino masses at the GUT scale
    in the normalization ${M_1}(GUT)$ = 1 and at the electroweak
    scale in the normalization ${M_1}(EW)$ = 1 
    for $F$ terms in different representations of $SU(5)$.
    These results are obtained by using 1-loop renormalization
    group equations.}
\renewcommand{\arraystretch}{1.0}
\end{table}
\noindent

\subsubsection{$SO(10)$}
\label{non-universal-SO(10)}

For the case of $SO(10)$, we have for the product of two 
adjoint ({\bf 45})-dimensional representations

\bea 
({\bf 45} \times {\bf 45})_{Symm}={\bf 1} 
\oplus {\bf 54} \oplus {\bf 210} \oplus {\bf 770}.
\label{symmetric_SO10}
\eea

\begin{table}[ht]
\begin{minipage}[b]{0.45\linewidth}
  \begin{tabular}{||c|c|ccc|ccc||}
   \hline 
   $SO(10)$ & $SU(5)$ & $M_1^G$ & $M_2^G$ & $M_3^G$ & 
    $M_1^{EW}$ & $M_2^{EW}$ & $M_3^{EW}$
    \\ \hline 
    {\bf 1} & {\bf 1} &1 &1 &1 &1 &2 &7.1\\[0.5 mm]
     {\bf 54} & {\bf 24} &1 &3 &-2 
     & 1 &6  &-14.3 \\[0.5 mm]
 {\bf 210} & {\bf 1}  &1 &1 &1 &1 &2 &7.1\\[0.5 mm]
           &  {\bf 24} &1 &3 &-2
     & 1 & 6  &-14.3 \\[0.5 mm]
           & {\bf 75} & 1 &-$\frac{3}{5}$ &-$\frac{1}{5}$ & 1 & -1.18 & -1.41 \\[0.5 mm]
  {\bf 770} & {\bf 1}  &1 &1 &1 &1 &2 &7.1\\[0.5 mm]
             &  {\bf 24} &1 &3 &-2
     & 1 &6  &-14.3 \\[0.5 mm]
           & {\bf 75} & 1 &-$\frac{3}{5}$ &-$\frac{1}{5}$ & 1 & -1.18 & -1.14 \\[0.5 mm]
           & {\bf 200} & 1 & $\frac{1}{5}$ &$\frac{1}{10}$ &1 & 0.4 & 0.71          
         \\ \hline
  \end{tabular}
   \caption{\label{tab2}Ratios of the gaugino masses at the GUT scale
    in the normalization ${M_1}(GUT)$ = 1 and at the electroweak
    scale in the normalization ${M_1}(EW)$ = 1 
    for $F$ terms in representations of $SU(5)\subset SO(10)$
    with the normal (nonflipped) embedding. These results have been 
    obtained at the 1-loop level.}
 \end{minipage}\qquad
\hspace{0.2cm}
\begin{minipage}[b]{0.49\linewidth}
\begin{center}
  \begin{tabular}{||c|c|ccc|ccc||}
   \hline 
   $SO(10)$ & $[SU(5)' \times U(1)]_{flipped}$ & $M_1^G$ & $M_2^G$ & $M_3^G$ & 
    $M_1^{EW}$ & $M_2^{EW}$ & $M_3^{EW}$
    \\ \hline 
    {\bf 1} & ({\bf 1},0) &1 &1 &1 &1 &2 &7.1\\ [0.5 mm]
     {\bf 54} & ({\bf 24},0) & 1 &3 &-2 
     & 1 &6  &-14.3 \\ [0.5 mm]
    {\bf 210} & ({\bf 1},0)  & 1 &-$\frac{5}{19}$ &-$\frac{5}{19}$ &1 &-0.52 &-1.85\\ [0.5 mm]
           &  ({\bf 24},0) & 1 &-$\frac{15}{7}$ &$\frac{10}{7}$ 
     & 1 &-4.2  &10 \\ [0.5 mm]
           & ({\bf 75},0) & 1 &-15 & -5 & 1 &-28 & -33.33 \\ [0.5 mm]
  {\bf 770} & ({\bf 1},0)  & 1 &$\frac{5}{77}$  &$\frac{5}{77}$ &1 &0.13 &0.46\\ [0.5 mm]
             &  ({\bf 24},0) & 1 &$\frac{15}{101}$ & -$\frac{10}{101}$  
     & 1 &0.3  &-0.70 \\ [0.5 mm]
           & ({\bf 75},0) & 1 &-15 &-5 & 1 &-28 &-33.3 \\ [0.5 mm]
           & ({\bf 200},0) & 1 &5 & $\frac{5}{2}$ &1 & 9.33 & 16.67          
         \\ \hline
  \end{tabular}
  \end{center}
  \caption{\label{tab3}Ratios of the gaugino masses at the GUT scale
    in the normalization ${M_1}(GUT)$ = 1 and at the electroweak
    scale in the normalization ${M_1}(EW)$ = 1 at the 1-loop level
    for $F$ terms in representations of flipped
    $SU(5)'\times U(1)$ $\subset SO(10)$.}
\end{minipage}

\vspace{1.0cm}

\begin{minipage}[b]{0.55\linewidth}
\begin{center}
   \begin{tabular}{||c|c|ccc|ccc||}
   \hline 
   $SO(10)$ & $ SU(4) \times SU(2)_R $ & 
                                          $M_1^G$ & $M_2^G$ & $M_3^G$ & 
    $M_1^{EW}$ & $M_2^{EW}$ & $M_3^{EW}$
    \\ \hline 
    {\bf 1} & ({\bf 1},{\bf 1}) &1 &1 &1 &1 &2 &7.1\\ [0.5 mm]
     {\bf 54} & ({\bf 1},{\bf 1}) & 1 &3 &2  
     & 1 &6  &-14.3 \\ [0.5 mm]
    {\bf 210} & ({\bf 1},{\bf 1})  & 1 &-$\frac{5}{3}$ &0 &1 &-3.35 &0\\ [0.5 mm]
           &  ({\bf 15},{\bf 1}) & 1 &0 &-$\frac{5}{4}$
     & 1 &0  &-9.09 \\ [0.5 mm]
           & ({\bf 15},{\bf 3}) & 1 & 0 & 0 &1  & 0 &0 \\ [0.5 mm]
  {\bf 770} & ({\bf 1},{\bf 1})  & 1 &$\frac{25}{19}$ &$\frac{10}{19}$&1 &2.6 &3.7\\ [0.5 mm]
             &  ({\bf 1},{\bf 5}) & 1 &0 &0  
     & 1 &0  &0 \\ [0.5 mm]
           & ({\bf 15},{\bf 3}) & 1 & 0 & 0 & 1 & 0 & 0 \\ [0.5 mm]
           & ({\bf 84},{\bf 1}) & 1 & 0 & $\frac{5}{32}$ &1 & 0 & 1.11          
         \\ \hline 
  \end{tabular}
  \end{center}
   \caption{\label{tab4}Ratios of the gaugino masses at the GUT scale
    in the normalization ${M_1}(GUT)$ = 1 and at the electroweak
    scale in the normalization ${M_1}(EW)$ = 1 at the 1-loop level
    for $F$ terms in representations of 
    $SU(4)\times SU(2)_L \times SU(2)_R \subset SO(10)$.} 
  \end{minipage}
  \end{table}

\noindent In Table~\ref{tab2} we have shown the gaugino mass parameters for the
different representations that arise in the symmetric product~(\ref{symmetric_SO10})
for the $SO(10)$ group. We note from Table~\ref{tab2} that the ratios of gaugino masses
for the different representations of $SO(10)$ in the symmetric product
(\ref{symmetric_SO10}) with the unflipped embedding 
$SU(5) \subset SO(10)$ are identical to
the corresponding gaugino mass ratios in Table~\ref{tab1} for 
the embedding of SM in $SU(5)$. Therefore, the input parameters and the 
resulting masses for the gaugino mass ratios in Table~\ref{tab2} for
$SO(10)$ are identical
to the corresponding  Tables~\ref{parEWSB24}, \ref{parEWSB75}, 
and \ref{parEWSB200} for $SU(5)$. 
There are two additional maximal power subgroups of $SO(10)$,
consistent with fermion content of the SM, apart from
$SU(5) \subset SO(10)$. We, thereforw, list in Tables~\ref{tab3}
and \ref{tab4}, the ratio of the gaugino mass parameters, both at
the GUT and electroweak scale, for different representations
that arise in the symmetric product of two adjoint representations
of $SO(10)$ with relevant embedding of these subgroups in $SO(10)$.

\section{NEUTRALINO MASS MATRIX, LAGRANGIAN, AND COUPLINGS}
\label{neut mass mat}

In this Appendix we recall the mixing matrix for the neutralinos
and the couplings that enter our calculations of the radiative neutralino 
cross section. We note that the neutralino mass matrix receives contribution
from MSSM superpotential term
\begin{eqnarray}
W_{\mathrm{MSSM}} & = & \mu H_1 H_2,
\label{WMSSM}
\end{eqnarray}
where $H_1$ and $H_2$ are the  Higgs doublet chiral superfields
with opposite hypercharge,
and $\mu$ is the supersymmetric Higgs(ino) mass parameter. In addition
to Eq.~(\ref{WMSSM}), the neutralino mass  matrix
receives contributions from the interactions between gauge and  matter
multiplets as well as contributions from the soft supersymmetry breaking 
masses for the $SU(2)_L$ and $U(1)_Y$  gauginos. 
Putting together all these contributions, the neutralino mass 
matrix, in the bino, wino, Higgsino basis 
$(-i\lambda', -i\lambda^3, \psi_{H_1}^1,
\psi_{H_2}^2)$, can be written as~\cite{Bartl:1989ms, Haber:1984rc}
\begin{eqnarray}
\label{mssmneut}
M_{\mathrm{MSSM}} =
\begin{pmatrix}
M_1 & 0   & - m_Z \sw \cos\beta & \phantom{-}m_Z\sw \sin\beta \\
0   & M_2 & \phantom{-} m_Z \cw \cos\beta  & -m_Z \cw\sin\beta \\
 - m_Z \sw \cos\beta &\phantom{-} m_Z \cw \cos\beta  & 0 & -\mu\\
\phantom{-}m_Z\sw \sin\beta& -m_Z \cw\sin\beta & -\mu & 0
\end{pmatrix},
\end{eqnarray}
where $M_1$ and $M_2$ are the $U(1)_Y$ and the $SU(2)_L$
soft gaugino mass parameters, respectively, and
$\tan\beta = v_2 /v_1$ is the ratio of the vacuum expectation
values of the neutral components of the two Higgs doublet 
fields $H_1$ and $H_2$, respectively. Furthermore,
$m_Z$ is the $Z$ boson mass, and $\theta_W$ is the
weak mixing angle. In our analyses we are considering all parameters 
in the neutralino mass matrix to be real. In this case it 
can be diagonalised by an orthogonal matrix. 
If one of the 
eigenvalues of $M_{\rm MSSM}$ is negative, then we can diagonalize
this matrix using a unitary matrix $N$, the neutralino 
mixing matrix, to get a positive semidefinite diagonal 
matrix~\cite{Haber:1984rc} 
with the neutralino masses $m_{\chi_i^0}~(i = 1, 2, 3, 4)$ in order of
increasing value: 
\begin{eqnarray}
\label{mssmdiag}
N^\ast M_{\mathrm{MSSM}} N^{-1} =   \mathrm{diag}\begin{pmatrix}m_{\chi_1^0}, 
& m_{\chi_2^0}, & m_{\chi_3^0}, & m_{\chi_4^0} \end{pmatrix}.
\end{eqnarray}

For the minimal supersymmetric standard model, the 
interaction Lagrangian of neutralinos, electrons, selectrons, and 
$Z$ bosons  is summarized by~\cite{Haber:1984rc}
\begin{eqnarray}
{\mathcal L} &=& (- \frac {\sqrt{2}e}{\cw} N_{11}^*)
                    \bar{f}_eP_L\tilde{\chi}^0_1\tilde{e}_R
                 + \frac{e}{\sqrt{2} \sw} (N_{12} + \tw N_{11}) 
                   \bar{f}_e P_R\tilde{\chi}^0_1\tilde{e}_L \nonumber \\
     & & + \frac{e}{4 \sw \cw} \left(|N_{13}|^2 - |N_{14}|^2\right)
           Z_\mu \bar{\tilde{\chi}}_1^0\gamma^\mu \gamma^5\tilde{\chi}_1^0
           \nonumber \\
          && + e Z_\mu \bar{f}_e \gamma^\mu 
             \big[ \frac{1}{\sw\cw}\left(\frac{1}{2} - \sw[2]\right) P_L 
                     - \tw  P_R\big] {f}_e + \mathrm{h. c.},
\label{mssmlagrangian}
\end{eqnarray} 
with the electron, selectrons, neutralino, and $Z$ boson fields denoted by
$f_e$, $\tilde{e}_{L,R}$, $\tilde{\chi}_1^0$, and $Z_\mu$, respectively,  
and $P_{R, L} = \frac{1}{2} \left(1 \pm \gamma^5\right)$. 
The interaction vertices 
arising from Eq.~(\ref{mssmlagrangian}) are summarized 
in Table~\ref{feynmandiag}.

\begin{table}[h!]
\begin{center}
\caption{Vertices corresponding to different terms in the interaction 
Lagrangian (\ref{mssmlagrangian}) for the MSSM. Here we have 
also shown the vertices for selectron-photon  and electron-photon
interactions~\cite{Basu:2007ys}.}
\vspace{5mm}
\begin{tabular}{lccccccccl}
\hline
\\
Vertex & & & & & & & & & Vertex factor\\
& & &  & &&& &&\\
\hline
& & & & & && &&\\
right~selectron - electron - neutralino 
& & & & & && && {$\frac {-i e \sqrt{2}}{\cw}N_{11}^* P_L$}\\
& & & & & && &&\\
left~selectron - electron - neutralino
& & & & & && &&{$\frac{i e}{\sqrt{2} \sw} (N_{12} + \tw N_{11}) P_R$}\\
& & & & & && &&\\
neutralino - $Z^0$ - neutralino
& & & & & && && {$ \frac{i e}{4 \sw \cw} 
           \left(|N_{13}|^2 - |N_{14}|^2\right) \gamma^\mu \gamma^5$}\\
& & & & & && && \\
electron - $Z^0$ - electron
& & & & & && && {$ i e \gamma^\mu \big[ \frac{1}{\sw\cw}\left(\frac{1}{2} 
               - \sw[2]\right) P_L - \tw  P_R\big] $} \\
& & & & & && &&\\
selectron - photon - selectron
& & & & & && && {$ i e (p_1 + p_2)^\mu$}\\
& & & & & && && \\
electron - photon - electron
& & & & & && && {$ i e \gamma^\mu$}\\
& & & & &  && && \\
\\
\hline
\bottomrule
\end{tabular}
\label{feynmandiag}
\end{center}
\end{table}


\end{document}